\pdfoutput=1

\documentclass{article}
\usepackage{ijcai24}

\usepackage{times}
\usepackage{soul}
\usepackage{url}
\usepackage[hidelinks]{hyperref}
\usepackage[utf8]{inputenc}
\usepackage[small]{caption}
\usepackage{graphicx}
\usepackage{amsmath}
\usepackage{amsthm}
\usepackage{booktabs}

\usepackage[linesnumbered,ruled,vlined]{algorithm2e}

\SetCommentSty{mycommfont}
\SetKwInput{KwData}{Input}
\SetKwInput{KwResult}{Output}

\newtheorem{example}{Example}
\newtheorem{theorem}{Theorem}
\newtheorem{definition}{Definition}
\newtheorem{lemma}{Lemma}
\newtheorem{observation}{Observation}
\newtheorem{fact}{Fact}
\newtheorem{corollary}{Corollary}
\newtheorem*{lemma*}{Lemma}
\newtheorem*{theorem*}{Theorem}

\usepackage[svgnames,table]{xcolor}
\usepackage[absolute,overlay]{textpos}
\usepackage{graphicx}
\usepackage{lipsum}
\usepackage{array}
\usepackage{afterpage}
\usepackage{blindtext}
\usepackage{pgffor}

\usepackage{pifont}

\usepackage{mathrsfs}

\usepackage[shortlabels]{enumitem} 
\usepackage{ltablex}
\usepackage{tabularx}

\usepackage{multirow}
\usepackage{multicol}
\usepackage{tcolorbox}

\usepackage{longtable}
\newcolumntype{N}{>{\small$}X<{$}}  
\newcolumntype{M}{>{\small$}l<{$}}  
\newcolumntype{L}{>{\small$}r<{$}}  
\newcolumntype{A}{>{\small}l<{}} 

\usepackage{stackengine}

\usepackage{appendix}

\usepackage{caption}
\usepackage{subcaption}

\usepackage{mathtools}
\usepackage{amsmath}
\usepackage{amssymb}
\usepackage{amsthm}
\usepackage{relsize}
\usepackage{thmtools}
\usepackage{thm-restate}
\usepackage{mathpartir}
\usepackage{listings}

\usepackage{tikz}

\usepackage{etoolbox}
\usepackage{environ}

\usepackage{hyperref}
\usepackage{cleveref}
\Crefname{Definition}{Definition}{Definitions}
\Crefname{Equation}{Equation}{Equations}
\Crefname{Figure}{Figure}{Figures}
\Crefname{Proposition}{Proposition}{Propositions}
\Crefname{Theorem}{Theorem}{Theorems}
\Crefname{Example}{Example}{Examples}
\Crefname{Corollary}{Corollary}{Corollaries}
\Crefname{Lemma}{Lemma}{Lemmata}
\Crefname{Section}{Section}{Sections}
\Crefname{Appendix}{Appendix}{Appendices}

\usepackage{fontawesome5}

\usepackage{scalerel}

\usepackage[activate={true,nocompatibility},final,tracking=true,
kerning=true,spacing=true,factor=1100,stretch=10,shrink=10]{microtype}
\SetTracking{encoding={*}, shape=sc}{40}

\usepackage{ellipsis}         
\usepackage[abbreviations,british]{foreign} 

\usepackage{etoolbox}

\usepackage{todonotes}
\usepackage{tikz}
\usepackage{xspace}

\usepackage{amssymb} 
\usepackage{amsmath}
\usepackage{amsthm}
\usepackage{mathtools}
\usepackage{mathrsfs}
\usepackage[bb=boondox]{mathalfa} 

\usepackage{tikz-cd}
\usetikzlibrary{matrix}

\makeatletter
\def\desclabel#1#2{\begingroup
\def\@currentlabel{#1}%
#1\label{#2}\endgroup
}
\makeatother

\usepackage{braket}

\usepackage[h]{esvect}

\usepackage[dash,dot]{dashundergaps}
\usepackage{float}
\usepackage{float}

\DeclareMathAlphabet{\mathdutchcal}{U}{dutchcal}{m}{n}
\SetMathAlphabet{\mathdutchcal}{bold}{U}{dutchcal}{b}{n}
\DeclareMathAlphabet{\mathdutchbcal}{U}{dutchcal}{b}{n}

\usetikzlibrary{arrows, decorations.markings, shapes, calc}

\usetikzlibrary{shadows}
\tikzset{
diagonal fill/.style 2 args={fill=#2, path picture={
\fill[#1, sharp corners] (path picture bounding box.south west) -|
                         (path picture bounding box.north east) -- cycle;}},
reversed diagonal fill/.style 2 args={fill=#2, path picture={
\fill[#1, sharp corners] (path picture bounding box.north west) |- 
                         (path picture bounding box.south east) -- cycle;}}
}

\usetikzlibrary{hobby,backgrounds,calc,trees}

\colorlet{jaune}{yellow!80!green}

\colorlet{vert}{green!45!black}

\colorlet{bleu}{blue!70!black}

\colorlet{rouge}{red!80!black}

\tikzstyle{minirond}=[draw, circle, minimum height=2mm, inner sep=0pt]
\tikzstyle{minirondd}=[draw, circle, minimum height=8mm, inner sep=0pt]
\tikzstyle{ptrond}=[draw, circle, minimum height=2.5mm]
\tikzstyle{medrond}=[draw, circle, minimum height=5mm]
\tikzstyle{rond}=[draw, circle, minimum height=7mm]

\tikzstyle{carre}=[draw,minimum width=6mm,minimum height=6mm]
\tikzstyle{medcarre}=[draw,minimum width=4mm,minimum height=4mm]
\tikzstyle{ptcarre}=[draw,minimum width=2.8mm,minimum height=2.8mm]
\tikzstyle{minicarre}=[draw,minimum width=1.5mm,minimum height=1.5mm,inner sep=0pt]
\tikzstyle{rouge}=[draw=red,fill=red!20!white]
\tikzstyle{vert}=[draw=green!80!black,fill=green!80!black!20!white]
\tikzstyle{jaune}=[draw=yellow!60!red,fill=yellow!60!red!30!white]
\tikzstyle{bleu}=[draw=blue,fill=blue!40!white]
\tikzstyle{gris}=[draw=black!80!white,fill=black!40!white]
\tikzstyle{rougef}=[draw=red,fill=red!60!white]
\tikzstyle{vertf}=[draw=green!80!black,fill=green!80!black!60!white]
\tikzstyle{jaunef}=[draw=yellow!80!black,fill=yellow!80!black!60!white]
\tikzstyle{bleuf}=[draw=blue,fill=blue!70!white]
\tikzstyle{rjaune}=[style=rond,style=jaune]
\tikzstyle{rbleu}=[style=rond,style=bleu]
\tikzstyle{rvert}=[style=rond,style=vert]
\tikzstyle{rrouge}=[style=rond,style=rouge]
\tikzstyle{rgris}=[style=rond,style=gris]
\tikzstyle{cjaune}=[style=carre,style=jaune]
\tikzstyle{cbleu}=[style=carre,style=bleu]
\tikzstyle{cvert}=[style=carre,style=vert]
\tikzstyle{crouge}=[style=carre,style=rouge]
\tikzstyle{cgris}=[style=carre,style=gris]
\tikzstyle{rjaunef}=[style=rond,style=jaunef]
\tikzstyle{rbleuf}=[style=rond,style=bleuf]
\tikzstyle{rvertf}=[style=rond,style=vertf]
\tikzstyle{rrougef}=[style=rond,style=rougef]

\newcommand{\tboxTloc}{\tboxT_{\textrm{loc}}}
\newcommand{\allTconssum}{\mathbf{S}_{\tboxT}}

\definecolor{ao(english)}{rgb}{0.0, 0.5, 0.0}

\newrobustcmd{\newnotion}[1]{\emph{#1}}
\newcommand{\deff}{\coloneqq}

\newcommand{\dland}{\sqcap}
\newcommand{\dlor}{\sqcup}

\newcommand{\dlsubseteq}{\sqsubseteq}
\newcommand{\Self}{\mathsf{Self}}
\newcommand{\bigdlor}{\bigsqcup}

\makeatletter
\providecommand{\bigdland}{%
  \mathop{%
    \mathpalette\@updown\bigsqcup
  }%
}
\newcommand*{\@updown}[2]{%
  \rotatebox[origin=c]{180}{$\m@th#1#2$}%
}
\makeatother

\makeatletter
\let\amsmath@bigm\bigm

\renewcommand{\bigm}[1]{%
  \ifcsname fenced@\string#1\endcsname
    \expandafter\@firstoftwo
  \else
    \expandafter\@secondoftwo
  \fi
  {\expandafter\amsmath@bigm\csname fenced@\string#1\endcsname}%
  {\amsmath@bigm#1}%
}

\newcommand{\DeclareFence}[2]{\@namedef{fenced@\string#1}{#2}}
\makeatother

\DeclareFence{\mid}{|}

\newcommand{\class}[1]{\mathscr{#1}}
\newcommand{\DL}[1]{\ensuremath{\mathcal{#1}}}  

\newcommand{\CQ}{\textrm{CQ}}   
\newcommand{\UCQ}{\textrm{UCQ}}

\newcommand{\ALC}{\DL{ALC}}                     
\newcommand{\ALCI}{\DL{ALCI}}                   

\newcommand{\ALCreg}{\DL{ALC}_{\textsf{reg}}}

\newcommand{\ALCHbselfreg}{\DL{ALCH}\textit{b}_{\textsf{reg}}^{\textsf{Self}}}

\newcommand{\Z}{\DL{Z}}                         
\newcommand{\ZIQ}{\DL{ZIQ}}                     
\newcommand{\ZOI}{\DL{ZOI}}                     
\newcommand{\ZOQ}{\DL{ZOQ}}                     
\newcommand{\ZOIQ}{\DL{ZOIQ}}                   

\newcommand{\SR}{\DL{SR}}                         

\newcommand{\abstrDL}{\DL{DL}}     

\newcommand{\complexityclass}[1]{\textsc{#1}} 
\newcommand{\NP}{\complexityclass{NP}} 
\newcommand{\PTime}{\complexityclass{PTime}} 
\newcommand{\coNExpTime}{\textrm{co}\complexityclass{NExpTime}} 
\newcommand{\ExpTime}{\complexityclass{ExpTime}} 
\newcommand{\NExpTime}{\complexityclass{NExpTime}} 
\newcommand{\lang}[1]{\mathbf{#1}}  
\newcommand{\Ilang}{\lang{N_I}}     
\newcommand{\Rlang}{\lang{N}_\lang{R}}     
\newcommand{\Clang}{\lang{N_C}}     

\newcommand{\query}[1]{\mathit{#1}}  
\newcommand{\queryq}{\query{q}}      
\newcommand{\match}[1]{#1}          
\newcommand{\matcheta}{\match{\eta}}  

\newcommand{\var}[1]{\mathit{#1}}   
\newcommand{\varx}{\var{x}}         
\newcommand{\vary}{\var{y}}         

\newcommand{\role}[1]{\mathit{#1}}      
\newcommand{\roler}{\role{r}}           
\newcommand{\roles}{\role{s}}           
\newcommand{\rolet}{\role{t}}           

\newcommand{\concepts}{\lang{C}}            

\newcommand{\concept}[1]{\mathrm{#1}}       
\newcommand{\conceptA}{\concept{A}}         
\newcommand{\conceptB}{\concept{B}}         
\newcommand{\conceptC}{\concept{C}}         
\newcommand{\conceptD}{\concept{D}}         

\newcommand{\indv}[1]{\textls[-85]{\texttt{#1}}}  
\newcommand{\indva}{\indv{a}}       
\newcommand{\indvb}{\indv{b}}       
\newcommand{\indvc}{\indv{c}}       
\newcommand{\indvo}{\indv{o}}       
\newcommand{\indvoo}{\indv{\"o}}       
\newcommand{\indvvo}{\indv{\v{o}}}       

\newcommand{\kb}[1]{\mathcal{#1}}   
\newcommand{\kbK}{\kb{K}}           
\newcommand{\abox}[1]{\mathcal{#1}} 
\newcommand{\aboxA}{\abox{A}}       
\newcommand{\aboxS}{\abox{S}}       
\newcommand{\tbox}[1]{\mathcal{#1}} 
\newcommand{\tboxT}{\tbox{T}}       
\newcommand{\ind}[1]{\mathsf{ind}{(#1)}} 
\newcommand{\indA}{\ind{\aboxA}}    
\newcommand{\indT}{\ind{\tboxT}}    
\newcommand{\indK}{\ind{\kbK}}    
\newcommand{\names}{\mathsf{N}}     

\newcommand{\inter}[1]{\mathcal{#1}}    
\newcommand{\interI}{\inter{I}}         
\newcommand{\interJ}{\inter{J}}         

\newcommand{\DeltaInter}[1]{\Delta^{#1}}                
\newcommand{\DeltaI}{\DeltaInter{\interI}}              

\newcommand{\cdotInter}[1]{\cdot^{#1}}          

\newcommand{\domelem}[1]{\mathrm{#1}}                           
\newcommand{\domelemc}{\domelem{c}}                             
\newcommand{\domelemd}{\domelem{d}}                             
\newcommand{\domeleme}{\domelem{e}}                             

\newcommand{\N}{\mathbb{N}}

\newcommand{\ZZ}{\mathbb{Z}}

\newcommand{\pathrho}{\rho}

\newcommand{\homo}[1]{\mathfrak{#1}}    
\newcommand{\homot}{\homo{t}}           

\newcommand{\interfwdunrav}[3]{{#1}^{#2}_{#3}}  
\newcommand{\Deltafwdunrav}[3]{\DeltaInter{\interfwdunrav{#1}{#2}{#3}}}  
\newcommand{\cdotfwdunrav}[3]{\cdotInter{\interfwdunrav{#1}{#2}{#3}}}    

\newrobustcmd{\interomegafwdunravI}[1]{\interfwdunrav{\interI}{\protect\vv{\omega}}{#1}}   
\newrobustcmd{\interomegafwdunravInames}{\interomegafwdunravI{\names}}  
\newrobustcmd{\interomegafwdunravIindA}{\interomegafwdunravI{\indA}}
\newrobustcmd{\interomegafwdunravIindK}{\interomegafwdunravI{\indK}}

\newrobustcmd{\DeltaomegafwdunravI}[1]{\Deltafwdunrav{\interI}{\protect\vv{\omega}}{#1}}   
\newrobustcmd{\DeltaomegafwdunravInames}{\DeltaomegafwdunravI{\names}}  
\newrobustcmd{\DeltaomegafwdunravIindA}{\DeltaomegafwdunravI{\indA}}

\newrobustcmd{\cdotomegafwdunravI}[1]{\cdotfwdunrav{\interI}{\protect\vv{\omega}}{#1}}     
\newrobustcmd{\cdotomegafwdunravInames}{\cdotomegafwdunravI{\names}}  
\newrobustcmd{\cdotomegafwdunravIindA}{\cdotomegafwdunravI{\indA}}

\newrobustcmd{\internfwdunravI}[1]{\interfwdunrav{\interI}{\protect\vv{n}}{#1}}   
\newrobustcmd{\internfwdunravInames}{\internfwdunravI{\names}}
\newrobustcmd{\internfwdunravIindK}{\internfwdunravI{\indK}}

\newrobustcmd{\internfwdunravJ}[1]{\interfwdunrav{\interJ}{\protect\vv{n}}{#1}}   
\newrobustcmd{\internfwdunravJnames}{\internfwdunravJ{\names}}
\newrobustcmd{\internfwdunravJindK}{\internfwdunravJ{\indK}}

\newrobustcmd{\DeltanfwdunravI}[1]{\Deltafwdunrav{\interI}{\protect\vv{n}}{#1}} 
\newrobustcmd{\DeltanfwdunravInames}{\DeltanfwdunravI{\names}}
\newrobustcmd{\DeltanfwdunravIindK}{\DeltanfwdunravI{\indK}}

\newrobustcmd{\cdotnfwdunravI}[1]{\cdotfwdunrav{\interI}{\protect\vv{n}}{#1}} 
\newrobustcmd{\cdotnfwdunravInames}{\cdotnfwdunravI{\names}}
\newrobustcmd{\cdotnfwdunravIindK}{\cdotnfwdunravI{\indK}}

\newcommand{\word}[1]{#1}                                       
\newcommand{\wordw}{\word{w}}                                   
\newcommand{\statesQ}{\mathtt{Q}}

\newcommand{\stateq}{\mathtt{q}}

\newcommand{\automatonA}{\mathdutchbcal{A}}
\newcommand{\automatonB}{\mathdutchbcal{B}}

\newcommand{\languageL}{\mathdutchbcal{L}}
\newcommand{\regexpR}{\mathdutchbcal{R}}

\newcommand{\rmN}{\mathrm{N}}

\definecolor{ForestGreen}{RGB}{34,139,34}
\definecolor{Salmon}{rgb}{1.0, 0.55, 0.41}
\definecolor{RawSienna}{rgb}{0.77, 0.12, 0.23}
\definecolor{MidnightBlue}{rgb}{0.0, 0.2, 0.4}
\definecolor{RedViolet}{rgb}{0.78, 0.08, 0.52}
\definecolor{TealBlue}{rgb}{0.21, 0.46, 0.53}
\definecolor{brandeisblue}{rgb}{0.0, 0.44, 1.0}

\usepackage{thmtools}
\usepackage[framemethod=TikZ]{mdframed}

\theoremstyle{definition}
\mdfdefinestyle{mdbluebox}{%
  roundcorner = 10pt,
  linewidth=1pt,
  skipabove=12pt,
  innerbottommargin=9pt,
  skipbelow=2pt,
  nobreak=true,
  linecolor=blue,
  backgroundcolor=TealBlue!8,
}
\declaretheoremstyle[
  headfont=\sffamily\bfseries\color{MidnightBlue},
  mdframed={style=mdbluebox},
  headpunct={\\[3pt]},
  postheadspace={0pt}
]{thmbluebox}

\mdfdefinestyle{mdredbox}{%
  linewidth=0.5pt,
  skipabove=12pt,
  frametitleaboveskip=5pt,
  frametitlebelowskip=0pt,
  skipbelow=2pt,
  frametitlefont=\bfseries,
  innertopmargin=4pt,
  innerbottommargin=8pt,
  nobreak=true,
  linecolor=RawSienna,
  backgroundcolor=Salmon!8,
}
\declaretheoremstyle[
  headfont=\bfseries\color{RawSienna},
  mdframed={style=mdredbox},
  headpunct={\\[3pt]},
  postheadspace={0pt},
]{thmredboxx}

\mdfdefinestyle{mdmyredbox}{%
  skipabove=8pt,
  linewidth=2pt,
  rightline=false,
  leftline=true,
  topline=false,
  bottomline=false,
  linecolor=RawSienna,
  backgroundcolor=Salmon!8,
}

\declaretheoremstyle[
  headfont=\bfseries\sffamily\color{ForestGreen!70!black},
  bodyfont=\normalfont,
  spaceabove=2pt,
  spacebelow=1pt,
  mdframed={style=mdmyredbox},
  headpunct={  },
]{thmmyredbox}

\mdfdefinestyle{mdgreenbox}{%
  skipabove=8pt,
  linewidth=2pt,
  rightline=false,
  leftline=true,
  topline=false,
  bottomline=false,
  linecolor=ForestGreen,
  backgroundcolor=ForestGreen!8,
}
\declaretheoremstyle[
  headfont=\bfseries\sffamily\color{ForestGreen!70!black},
  bodyfont=\normalfont,
  spaceabove=2pt,
  spacebelow=1pt,
  mdframed={style=mdgreenbox},
  headpunct={ },
]{thmgreenbox}

\declaretheoremstyle[
  headfont=\bfseries\sffamily\color{RawSienna!80!black},
  bodyfont=\normalfont,
  spaceabove=2pt,
  spacebelow=1pt,
  mdframed={style=mdmyredbox},
  headpunct={ },
]{thmredbox}

\declaretheoremstyle[
  headfont=\bfseries\sffamily\color{ForestGreen!70!black},
  bodyfont=\normalfont,
  spaceabove=2pt,
  spacebelow=1pt,
  mdframed={style=mdgreenbox},
  headpunct={},
]{thmgreenbox*}

\mdfdefinestyle{mdblackbox}{%
  skipabove=8pt,
  linewidth=3pt,
  rightline=false,
  leftline=true,
  topline=false,
  bottomline=false,
  linecolor=black,
  backgroundcolor=RedViolet!8!gray!8,
}
\declaretheoremstyle[
  headfont=\bfseries,
  bodyfont=\normalfont\small,
  spaceabove=0pt,
  spacebelow=0pt,
  mdframed={style=mdblackbox}
]{thmblackbox}

\mdfdefinestyle{mdbluebox}{%
  skipabove=8pt,
  linewidth=2pt,
  rightline=false,
  leftline=true,
  topline=false,
  bottomline=false,
  linecolor=brandeisblue,
  backgroundcolor=brandeisblue!8,
}
\declaretheoremstyle[
  headfont=\bfseries\sffamily\color{brandeisblue!70!black},
  bodyfont=\normalfont,
  spaceabove=2pt,
  spacebelow=1pt,
  mdframed={style=mdbluebox},
  headpunct={ },
]{thmbluebox}

\definecolor{ForestGreen}{RGB}{34,139,34}
\newcommand{\vocab}[1]{\textbf{\color{ForestGreen!70!black}#1}}

\newcommand{\IlangA}{\lang{N}_\lang{I}^{\aboxA}}    
\newcommand{\IlangT}{\lang{N}_\lang{I}^{\tboxT}}

\newcommand{\myqed}{\hfill{$\blacktriangleleft$}}
\newcommand{\Rlangsimpl}{\Rlang^{\text{sp}}}
\newcommand{\deadend}{{\scaleobj{0.75}{\text{\faSkull}}}}
\usepackage{halloweenmath}
\newcommand{\ghost}{{\scaleobj{0.85}{\mathghost}}}
\definecolor{cobalt}{rgb}{0.0, 0.28, 0.67}

\usepackage{stfloats}
\usepackage{algorithmic}
\usepackage[switch]{lineno}

\usepackage{stmaryrd}

\title{Data Complexity in Expressive Description Logics With Path Expressions}

\author{
    Bartosz Bednarczyk
    \affiliations
    Computational Logic Group, TU Dresden, Germany\\
    Institute of Computer Science, University of Wrocław, Poland
}

\begin{document}
\maketitle
\begin{abstract}    
    We investigate the data complexity of the satisfiability problem for the very expressive description logic $\ZOIQ$ (a.k.a. $\ALCHbselfreg\mathcal{OIQ}$) over quasi-forests and establish its $\NP$-completeness.
    This completes the data complexity landscape for decidable fragments of $\ZOIQ$, and reproves known results on decidable fragments~of~OWL2~($\SR$~family).
    Using the same technique, we establish $\textrm{co}\complexityclass{NExpTime}$-completeness (w.r.t. the combined complexity) of the entailment problem of rooted queries~in~$\ZIQ$.
\end{abstract}


\section{Introduction}
Formal ontologies are essential in nowadays approaches to safe and trustworthy symbolic AI, serving as the backbone of the Semantic Web, peer-to-peer data management, and information integration.
Such applications often resort to managing and reasoning about graph-structure data. 
A premier choice for the reasoning framework is then the well-established family of logical formalisms known as \emph{description logics} (DLs) \cite{dlbook} that underpin the logical core of OWL~2 by the W3C~\cite{OWL2Primer}.
Among the plethora of various features available in extensions of the basic DL called $\ALC$, an especially prominent one is $\cdot_{\mathsf{reg}}$, supported by the popular $\Z$-family of DLs~\cite{CalvaneseEO09}.
Using $\cdot_{\mathsf{reg}}$, one can specify regular path constraints and hence, allow the user to navigate the underlying graph data-structure.
In recent years, many extensions of $\ALCreg$ for ontology-engineering were proposed~\cite{BienvenuCOS14,CalvaneseOS16,OrtizS12}, and the complexity of their reasoning problems is mostly well-understood:~Calvanese~\emph{et al.} \shortcite{CalvaneseEO09,CalvaneseOS16}; Bednarczyk \emph{et al.}~\shortcite{BednarczykR19,BednarczykK22}. 

Vardi~\shortcite{Vardi82} observed that measuring the user's data and the background ontology equally is not realistic, as the data tends to be huge in comparison to the~ontology.
This gave~rise to the notion of \emph{data complexity}: the ontology (TBox) is fixed upfront and only the user's data (ABox) varies.
The \emph{satisfiability} problem for DLs is usually $\NP$-complete w.r.t. the data complexity, including the two-variable counting logic~\cite{PrattHartmann09} (encoding DLs up to $\ALC\mathcal{BIOQ}\Self$) as well as $\mathcal{SROIQ}$~\cite{Kazakov08}, the logical core of OWL2. 
Regarding DLs with path expressions, $\NP$-completeness of $\ALC\mathcal{I}_{\textsf{reg}}^{\Self}$~\cite{JungLM018} was established only recently.

\subsection{Our Main Contribution and a Proof Overview}
We study the data complexity of the satisfiability problem for decidable sublogics of $\ZOIQ$ (a.k.a. $\ALCHbselfreg\mathcal{OIQ}$), and establish $\NP$-completeness of $\ZIQ$, $\ZOQ$, and~$\ZOI$.  
As all the mentioned DLs possess the \emph{quasi-forest model property} (\ie every satisfiable knowledge-base (KB) has a forest-like model), for the uniformity of our approach we focus on the satisfiability of $\ZOIQ$ over quasi-forests.
Calvanese et al.~\shortcite{CalvaneseEO09} proved that quasi-forest-satisfiability of $\ZOIQ$ is $\ExpTime$-complete w.r.t. the combined complexity. Unfortunately, their approach is automata-based and relies on an internalisation of ABoxes inside $\ZOIQ$-concepts, and thus cannot be used to infer tight bounds w.r.t. the data complexity.

We employ the aforementioned algorithm of Calvanese et al. as a black~box and design a novel algorithm for quasi-forest-satisfiability of $\ZOIQ$-KBs.
In our approach we construct a quasi-forest model in two steps, \ie we construct its root part (the \emph{clearing}) separately from its subtrees. 
Our algorithm first pre-computes (an exponential w.r.t. the size of the TBox but of constant size if the TBox is fixed) set of quasi-forest-satisfiable $\ZOIQ$-concepts that indicate possible subtrees that can be ``plugged in'' to the clearing of the intended model. 
Then it guesses (in $\NP$) the intended clearing and verifies its consistency in $\PTime$ based on the pre-computed concepts and roles.
For the feasibility of our  ``modular construction'' a lot of bookkeeping needs to be done.
Most importantly, certain \emph{decorations} are employed to decide the satisfaction of \emph{automata concepts} and \emph{number restrictions} by elements in an incomplete and fragmented forest.
The first type of decorations, given an automaton $\automatonA$, aggregates information about existing paths realising $\automatonA$ and starting at one of the roots of the intended model. 
As a single such path may visit several subtrees, we cut such paths into relevant pieces and summarise them by means of ``shortcut'' roles and $\ZOIQ$-concepts describing paths fully contained inside a single subtree.
The second type of decorations ``localise'' counting in the presence of nominals, as the nominals may have successors outside their own subtree and the clearing. 
These two ``small tricks'', obfuscated by various technical difficulties, are the core ideas underlining our proof. 
We conclude the paper by presenting how our algorithm can be adapted to the entailment problem of rooted queries over $\ZIQ$. The key idea here is to guess an ``initial segment'' of a quasi-forest with no query match, and check if it can be extended to a full model~of~the~input~KB.

\section{Preliminaries}\label{sec:preliminaries}

We assume familiarity with description logics (DLs)~\cite{dlbook}, formal languages, and complexity~\cite{sipser13}.  
We fix countably-infinite pairwise-disjoint sets $\Ilang, \Clang, \Rlang$ of \emph{individual}, \emph{concept}, and \emph{role} \emph{names}.
An \emph{interpretation} $\interI \deff (\DeltaI, \cdot^{\interI})$ consists of a non-empty domain $\DeltaI$ and a function~$\cdot^{\interI}$ that maps $\indva \in \Ilang$ to $\indva^{\interI} \in \DeltaI$, $\conceptA \in \Clang$ to $\conceptA^{\interI} \subseteq \DeltaI$, and $\roler \in \Rlang$~to~$\roler^{\interI} \subseteq \DeltaI \times \DeltaI$.
The set of \emph{simple roles}~$\Rlangsimpl$~is defined with the grammar $\roles \Coloneqq \roler \in \Rlang \mid \roles^- \mid \roles {\cap} \roles \mid \roles {\cup} \roles \mid \roles {\setminus} \roles$.
Simple roles are interpreted as expected by invoking the underlying set-theoretic operations, \eg $\interI$ interprets $(\roler \cup (\roles \setminus \rolet) )^-$ as the inverse of $(\roler^{\interI} \cup (\roles^{\interI} \setminus \rolet^{\interI}))$. 
The superscripts $^*$ and $^+$ denote the Kleene's star and plus, while $\cdot$ denotes concatenation.
A~\emph{path} in~$\interI$ is a word $\pathrho \deff \pathrho_1 \ldots \pathrho_{|\pathrho|}$ in~$(\DeltaI)^+$ such that for all~$i < |\pathrho|$ there is a role name $\roler \in \Rlangsimpl$ with $(\pathrho_i, \pathrho_{i{+}1}) \in \roler^{\interI}$.

\paragraph*{Automata.} We use nondeterministic automata (NFAs). For states $\stateq$ and~$\stateq'$ of an NFA~$\automatonA$, $\automatonA_{\stateq,\stateq'}$ denotes the corresponding NFA with the initial (resp. final) state switched to $\stateq$ (resp. $\stateq'$). NFAs use roles from $\Rlangsimpl$ and \emph{tests}~$\conceptC?$ for concepts $\conceptC$ as alphabet.
A path $\pathrho$ \emph{realises} $\automatonA$~($\pathrho \models \automatonA$, $\pathrho$ is an $\automatonA$-path)~if $\automatonA$ accepts some word $\wordw_1\roler_1\ldots\wordw_{|\pathrho|{-}1}\roler_{|\pathrho|{-}1}\wordw_{|\pathrho|}$, where $\roler_i \in \Rlangsimpl$ and $\wordw_i$ are (possibly empty) sequences of tests, satisfying $(\pathrho_i, \pathrho_{i{+}1}) \in \roler_i^{\interI}$  and $\pathrho_i \in \conceptC^{\interI}$ for all $i \leq |\pathrho|$ and tests~$\conceptC?$~in~$\wordw_i$.

\paragraph*{Expressive DLs.}
The DL $\ZOIQ$~\cite{CalvaneseEO09} is the main object of our study.
For all $\conceptA \in \Clang$, $\roles \in \Rlangsimpl$, $\indvo \in \Ilang$, $n \in \N$ and NFA $\automatonA$, we define $\ZOIQ$-concepts~$\conceptC$~with:
\begin{align*}
  \bot \mid \conceptA \mid \{ \indvo \} \mid \neg \conceptC \mid \conceptC \dland \conceptC \mid \exists{\roles}.\conceptC \mid ({\geq}{n}\ \roles).\conceptC \mid \exists{\roles}.\Self \mid \exists{\automatonA}.\conceptC
\end{align*}
We denote $\top \deff \neg\bot$, $\conceptC \dlor \conceptC' \deff \neg(\neg\conceptC \dland \neg\conceptC')$, $({\leq}{n}\; \roles).\conceptC \deff \neg({\geq}{n{+}1}\; \roles).\neg\conceptC$, $({=}{n}\; \roles).\conceptC \deff ({\leq}{n}\; \roles).\conceptC \dland ({\geq}{n}\; \roles).\conceptC$, and $\forall{\heartsuit}.\conceptC \deff \neg\exists{\heartsuit}.\neg\conceptC$, where $\heartsuit$ is a simple role or an~NFA.
\vspace{-0.8em}
\begin{table}[H]
  \begin{center}
  \scalebox{.95}{
          \begin{tabular}{@{}l@{\ \ \ }c@{\ \ \ }r@{}}
              \hline\\[-2ex]
              Name & Syntax & Semantics: $d \in \conceptC^{\interI}$ if  \\\hline\\[-2ex]
              top/bottom & $\top/ \bot$ & always/never\\ 
              nominal   & $\{ \indvo \}$ &  $\domelemd = \indvo^{\interI}$ \\
              negation & $\neg \conceptC$ & $\domelemd \not\in \conceptC^{\interI}$\\
              intersection & $\conceptC_1 \dland \conceptC_2$ & $\domelemd \in \conceptC_1^{\interI}$ and  $\domelemd \in \conceptC_2^{\interI}$\\
              existential restriction & $\exists{\roles}.\conceptC$ & $\exists\domeleme\left( \domeleme \in \conceptC^{\interI} \land (\domelemd, \domeleme) \in \roles^{\interI}\right)$\\
              number restriction & $({\geq}{n}\; \roles).\conceptC$ & $|\{ \domeleme \in \conceptC^{\interI}\mid (\domelemd,\domeleme) {\in} \roles^{\interI}\}| \ge n$\\
              $\Self$ concept & $\exists \roles.\Self$ & $(\domelemd,\domelemd)\in \roles^{\interI}$\\
              automaton concept & $\exists{\automatonA}.\conceptC$ & 
              $\exists{\text{path}}$ $\domelemd{\cdot}\pathrho{\cdot}\domeleme \models \automatonA$ \& $\domeleme \in \conceptC^{\interI}$
              \\\hline\\[-2ex]
      \end{tabular}}
    \end{center}
  \end{table}
  \vspace{-1.65em}
  \noindent We define the DLs $\ZOI$, $\ZIQ$, and $\ZOQ$, respectively, by dropping (a) number restrictions, (b) nominals, and (c) role inverses from the~syntax of $\ZOIQ$. W.l.o.g.\@ we allow for regular expression (with tests) in place of NFAs in above~syntax.

  A \emph{$\ZOIQ$-KB} $\kbK\deff (\aboxA, \tboxT)$ is composed of two sets of \emph{axioms}, an \emph{ABox} $\aboxA$ and a \emph{TBox}~$\tboxT$. An interpretation $\interI$ is a \emph{model} of $\kbK$ ($\interI \models \kbK$) if it satisfies all the axioms of $\kbK$. The content of ABoxes and TBoxes, as well as the notion of their satisfaction is defined below, assuming that $\indva, \indvb \in \Ilang$, $\conceptA \in \Clang$, $\roler \in \Rlang$, $\roles, \rolet \in \Rlangsimpl$, and $\conceptC, \conceptD$ are $\ZOIQ$-concepts.
  \vspace{-1.6em}
  \begin{table}[H]
    \centering
        \begin{tabular}{ l l }
            \hline\\[-2ex]
            Axiom  $\alpha$ & $\interI \models\alpha$, if \\ \hline\\[-2ex] 
            $\conceptC \dlsubseteq \conceptD, \roles \subseteq \rolet$ & $\conceptC^\interI \subseteq \conceptD^\interI, \roles^{\interI} \subseteq \rolet^{\interI}$\hspace{10.4ex} \mbox{TBox } $\tboxT$\\\hline\\[-2ex]
            $\conceptA(\indva),\ \roler(\indva, \indvb)$ & $\indva^\interI \in \conceptA^\interI$,\ $(\indva^\interI ,\indvb^\interI ) \in \roler^\interI $ \; \; \; \; \hfill\mbox{ABox }$ \aboxA$\\            
            $\indva \approx \indvb, \ \neg \alpha$ & $\indva^\interI = \indvb^\interI, \ \interI \not\models \alpha$\\ \hline
        \end{tabular}
\end{table}

Note that ABoxes do not involve complex concepts. Otherwise, data complexity of $\ALCreg$ is already $\ExpTime$-hard.

\paragraph*{} \noindent \hspace{-1em}A TBox is in \emph{Scott's normal form} if its axioms have~the~form:
  \begin{align*}
    \vspace{-0.5em}
    \conceptA \equiv \conceptB, \; \; \; \conceptA \equiv \neg \conceptB, \; \; \; \conceptA \equiv \{ \indvo \}, \; \; \; \conceptA \equiv \conceptB \dland \conceptB',  \; \; \; \roler = \roles,\\
    \vspace{-0.2em}
    \conceptA \equiv \exists{\automatonA_{\stateq,\stateq'}.\top}, \; \; \; \conceptA \equiv \exists{\roler}.\Self, \; \; \; \conceptA \equiv ({\geq}{n}\; \roler).\top
    \vspace{-0.5em}
  \end{align*}
  for  $\conceptA, \conceptB, \conceptB' \in \Clang \cup \{ \top, \bot \}$, 
  $\indvo \in \Ilang$,
  $\roler \in \Rlang$, 
  $\roles \in \Rlangsimpl$,
  $n \in \N$, 
  and NFAs $\automatonA_{\stateq,\stateq'}$.
  Here $\interI \models \conceptC\equiv\conceptD$ iff $\conceptC^{\interI}=\conceptD^{\interI}$.~W.l.o.g. we focus only on KBs with TBoxes~as~above  [Appendix B].

\paragraph*{Queries.} 
We focus on \emph{conjunctive queries} ({\CQ}s) defined by the grammar: $\queryq \Coloneqq \conceptA(\varx) \ \mid \  \roler(\varx, \vary) \ \mid \ \queryq \land \queryq$, where $\conceptA \in \Clang$, $\roler \in \Rlang$, and $\varx, \vary$ being either~variables or individual names.
An interpretation~$\interI$ \emph{satisfies} a query~$\queryq$ ($\interI \models \queryq$), if there is an assignment $\matcheta$  (a \emph{match}, defined as expected), mapping variables to domain elements and individual names $\indva \in \Ilang$ to the corresponding $\indva^{\interI}$, under which $\queryq$ evaluates to true.
A~{\CQ}~is \emph{rooted} if it has at least one individual name and its query graph is connected.
A KB  \emph{entails} $\queryq$ if every of its models satisfies~$\queryq$.

\paragraph*{Problems.}
Let $\abstrDL$ be a logic, and $\class{Q}$ be a class of queries.
In the \emph{satisfiability problem} we ask if a given $\abstrDL$-KB $\kbK \deff (\aboxA, \tboxT)$ has a model. 
In the \emph{(rooted) query entailment problem} we ask if a given $\abstrDL$-KB $\kbK$ entails a (rooted) query $\queryq \in \class{Q}$.
We distinguish two ways of measuring the size of the input, which gives rise to \emph{combined} and \emph{data} \emph{complexity}.
In the first case all the components $\aboxA, \tboxT$ and $\queryq$ equally contribute to the size of the input. 
In the second case, $\tboxT$ and $\queryq$ are assumed to be fixed beforehand, and thus their size is treated as constant.

\paragraph*{Forests.} 
We adapt the notion of \emph{quasi-forests}~\cite{CalvaneseEO09} under the standard set-theoretic reconstruction of the notion of an $\N$-forest as a prefix-closed subset~of~$\N^+$ without $\varepsilon$.
Our notion is stricter than the original definition by Calvanese~\emph{et al.} but the differences are negligible [Appendix~C]. 

\begin{definition}\label{def:quasi-forest}
Let $\IlangA$ and $\IlangT$ be finite subsets of $\Ilang$, $\concept{Root} \in \Clang$, and $\role{child}, \role{edge}, \role{id} \in \Rlang$.
An interpretation $\interI$ is an \vocab{$(\IlangA, \IlangT)$-quasi-forest} if its domain $\DeltaI$ is an  $\N$-forest,
\begin{align*}
\concept{Root}^{\interI} & = & \DeltaI {\cap} \N = \{ \indva^{\interI} \mid \indva {\in} \Ilang \} = \{ \indva^{\interI} \mid \indva {\in} (\IlangA {\cup} \IlangT) \},\\
\role{child}^{\interI} & = & \{ (\domelemd, \domelemd {\cdot} n) \mid \domelemd, \domelemd {\cdot} n \in \DeltaI, n \in \N \},\\
\role{edge}^{\interI} & = &  \textstyle\bigcup_{\roler \in \Rlang} \roler^{\interI} \cup (\roler^-)^\interI,\\
\role{id}^{\interI} & = & \{ (\domelemd, \domelemd) \mid \domelemd \in \DeltaI \},
\end{align*}
\vspace{-0.01em}
and for all roles $\roler^{\interI}$ and all pairs $(\domelemd, \domeleme) \in \roler^{\interI}$ at least one of the  conditions hold:
(i) both $\domelemd$ and $\domeleme$ belong to $\concept{Root}^\interI$,
(ii) one of $\domelemd$ or $\domeleme$ is equal to $\indvo^{\interI}$ for some name $\indvo \in \IlangT$,
(iii) $(\domelemd,\domeleme) \in \role{id}^{\interI} \cup \role{child}^{\interI} \cup (\role{child}^-)^{\interI}$.
For convenience, we refer to pairs $(\domelemd, \domeleme) \in \roler^{\interI}$ satisfying the second condition as \emph{backlinks}, and the ones satisfying $(\domelemd, \domeleme) \in \role{id}^{\interI}$ as \emph{self-loops}.
The \emph{clearing} of $\interI$ is the restriction of $\interI$ to $\concept{Root}^{\interI}$.
A quasi-forest is \emph{$\rmN$-bounded} if every domain element has at most $\rmN$ $\role{child}$-successors. 
\myqed
\end{definition}

The names from $\IlangT$ are dubbed \emph{nominals}, and will be usually denoted with decorated letters $\indvo$.
Their interpretations are usually referred as \emph{nominal roots}.
Throughout the paper we employ suitable notions~from~graph theory such as \emph{node}, \emph{root}, \emph{child}, \emph{parent}, or \emph{descendant}, defined as expected in accordance with $\DeltaI$, $\role{child}^{\interI}$ and $\concept{Root}^{\interI}$. For instance, $\domelemd$ is a descendant of $\domelemc$ whenever $(\domelemc, \domelemd) \in (\role{child}^{\interI})^+$ holds. Consult \Cref{ex:quasi-forest}.

\begin{example}\label{ex:quasi-forest}
  An example $3$-bounded $(\{ \indva, \indvb \}, \{ \indvo, \indvoo, \indvvo \} )$-quasi-forest $\interI$ is depicted below. For readability we have omitted the interpretations of $\concept{Root}, \role{child}$, $\role{id}$, and $\role{edge}$.
  
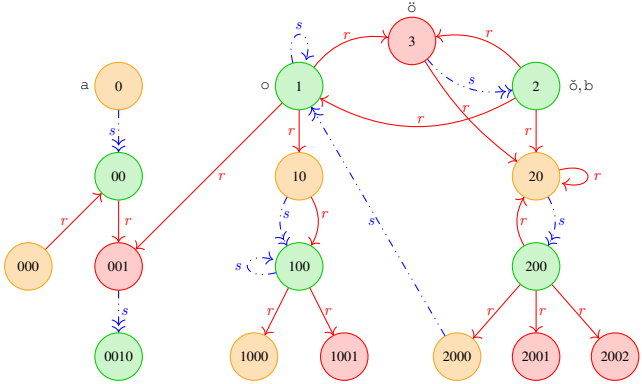
\begin{figure}[H]
    \centering
    \vspace{-1em}
    \begin{tikzpicture}[scale=0.6, transform shape]
      \draw (0,0) node[ptrond, jaune, label=center:\small{0}] (A1) {\phantom{0000}};
      \node[] at (-0.75, 0) {\( \indva \)};
  
      \draw (0,-2) node[ptrond, vert, label=center:\small{00}] (A12) {\phantom{0000}};
      \draw (-2,-4) node[ptrond, jaune, label=center:\small{000}] (A121) {\phantom{0000}};
      \draw (0,-4) node[ptrond, rouge, label=center:\small{001}] (A124) {\phantom{0000}};
      \draw (0, -6) node[ptrond, vert, label=center:\small{0010}] (A1243) {\phantom{0000}};
  
      \path[->] (A121) edge [red] node[xshift=-6] {\(\roler \)} (A12);
      \path[->] (A12) edge [red] node[xshift=6] {\(\roler \)} (A124);
      \path[->>,dashdotdotted] (A124) edge [blue] node[xshift=4] {\(\roles \)} (A1243);
      \path[->>,dashdotdotted] (A1) edge [blue] node[xshift=-4] {\(\roles \)} (A12);
  
      \draw (4, 0) node[ptrond, vert, label=center:\small{1}] (A2) {\phantom{0000}};
      \node[] at (3.25, 0) {\(\indvo \)};
  
      \draw (4, -2) node[ptrond, jaune, label=center:\small{10}] (A21) {\phantom{0000}};
      \draw (4, -4) node[ptrond, vert, label=center:\small{100}] (A212) {\phantom{0000}};
      \draw (3, -6) node[ptrond, jaune, label=center:\small{1000}] (A2121) {\phantom{0000}};
      \draw (5, -6) node[ptrond, rouge, label=center:\small{1001}] (A2124) {\phantom{0000}};
  
      \path[->] (A2) edge [red] node[xshift=-5] {\(\roler \)} (A21);
  
      \path[->] (A21) edge [bend left=30, red] node[yshift=4.5,xshift=4] {\(\roler \)} (A212);
      \path[->>,dashdotdotted] (A21) edge [bend right=30, blue] node[xshift=4, yshift=4] {\(\roles \)} (A212);
  
      \path[->] (A212) edge [red] node[xshift=-4] {\(\roler \)} (A2121);
      \path[->] (A212) edge [red] node[xshift=4] {\(\roler \)} (A2124);
  
      \draw (9.25, 0) node[ptrond, vert, label=center:\small{2}] (A3) {\phantom{0000}};
      \node[] at (10.25, 0) {\(\indvvo, \indvb \)};
  
      \draw (9.25, -2) node[ptrond, jaune, label=center:\small{20}] (A31) {\phantom{0000}};
      \draw (9.25, -4) node[ptrond, vert, label=center:\small{200}] (A312) {\phantom{0000}};
      \draw (7.5, -6) node[ptrond, jaune, label=center:\small{2000}] (A3121) {\phantom{0000}};
      \draw (9.25, -6) node[ptrond, rouge, label=center:\small{2001}] (A3124) {\phantom{0000}};
      \draw (11, -6) node[ptrond, rouge, label=center:\small{2002}] (A3126) {\phantom{0000}};

      \path[->] (A3) edge [red] node[xshift=-5] {\(\roler \)} (A31);
      \path[->>,dashdotdotted] (A31) edge [bend left=30, blue] node[xshift=4] {\(\roles \)} (A312);
      \path[->] (A312) edge [bend left=30, red] node[xshift=4] {\(\roler \)} (A31);
      \path[->] (A312) edge [red] node[xshift=-4] {\(\roler \)} (A3121);
      \path[->] (A312) edge [red] node[xshift=4] {\(\roler \)} (A3124);
      \path[->] (A312) edge [red] node[xshift=4] {\(\roler \)} (A3126);
  
      \draw (6.5, 1) node[ptrond, rouge, label=center:\small{3}] (A4) {\phantom{0000}};
      \node[] at (6.5, 1.75) {\(\indvoo \)};
  
      \path[->] (A31) edge [loop right,red] node {$\roler$} ();
      \path[->>,dashdotdotted] (A212) edge [loop left,blue] node {$\roles$} ();
  
      \path[->] (A2) edge [bend left=30, red] node[yshift=4] {\(\roler \)} (A4);
      \path[->] (A3) edge [bend left=30, red] node[yshift=4.5,xshift=2] {\(\roler \)} (A2);
      \path[->] (A3) edge [bend right=30, red] node[xshift=4, yshift=4] {\(\roler \)} (A4);
      \path[->>,dashdotdotted] (A4) edge [bend right=30, blue] node[xshift=4,yshift=4] {\(\roles \)} (A3);
      \path[->>,dashdotdotted] (A2) edge [loop above,blue] node {$\roles$} ();

  
      \path[->>,dashdotdotted] (A3121) edge [blue] node[xshift=-4] {\(\roles \)} (A2);
      \path[->] (A4) edge [bend right=10, red] node[xshift=0, yshift=5] {\(\roler \)} (A31);
      \path[->] (A2) edge [red] node[xshift=8] {\(\roler \)} (A124);
  
  \end{tikzpicture}
  \end{figure}  
  \noindent The element $1$ has a unique child $10$.  
  The roots of $\interI$ are $0$, $1$, $2$, and $3$. The nominal roots are $1$,$2$, and $3$.
   The pairs $(1,001)$, $(2000,1)$, and $(3,20)$ are example backlinks.\myqed
\end{example}

A \emph{quasi-forest model} of a $\ZOIQ$-KB $\kbK \deff (\aboxA, \tboxT)$ is an $(\indA, \ind{\tboxT})$-quasi-forest satisfying $\kbK$, where $\indA$ and $\ind{\tboxT}$ are sets of all individual names from $\aboxA$ and~$\tboxT$.
The~results by Calvanese \emph{et al.}~\shortcite[Prop.~3.3]{CalvaneseEO09} and Ortiz~\shortcite[L.~3.4.1, Thm.~3.4.2]{OrtizPhD10} advocate the use of quasi-forest models. 

\begin{lemma}\label{lemma:quasi-forest-models-by-Ortiz-et-al}
  Let $\kbK {\deff} (\aboxA, \tboxT)$ be a KB of $\ZOQ$, $\ZOI$, or $\ZIQ$. 
  Then is an integer $\rmN$ (exponential w.r.t. $|\tboxT|$) such~that:\\
  (I) $\kbK$ is satisfiable iff $\kbK$ has an $\rmN$-bounded quasi-forest model, and (II) for all (unions of) {\CQ}s $\queryq$, we have $\kbK \not\models \queryq$ iff there exists an $\rmN$-bounded quasi-forest model of $\kbK$ violating $\queryq$.\myqed
\end{lemma}

\noindent The quasi-forest (counter)models of $\ZOIQ$-KBs  are \emph{canonical} if they are $\rmN$-bounded for the $\rmN$ guaranteed by \Cref{lemma:quasi-forest-models-by-Ortiz-et-al}.
Hence, for the deciding the satisfiability and query entailment over $\ZIQ$, $\ZOQ$, and $\ZOI$, only the class of canonical~quasi-forest~models~is~relevant.
We say that a $\ZOIQ$-KB is \emph{quasi-forest satisfiable} whenever it has a canonical quasi-forest model.
The quasi-forest satisfiability problem is defined accordingly.
We have~\cite[L~3.4.1, Thm.~3.4.2]{OrtizPhD10} that:

\begin{lemma}\label{lemma:deciding-ZOIQ-ExpTime-complete}
  The quasi-forest satisfiability problem for $\ZOIQ$-KBs is $\ExpTime$-complete (w.r.t. the combined complexity). 
  In particular, the satisfiability of $\ZIQ$, $\ZOQ$, and $\ZOI$-KBs is $\ExpTime$-complete (w.r.t. the combined complexity).\myqed
\end{lemma} 

\section{Basic Paths in Quasi-Forests}\label{sec:basic-paths-in-quasi-forests}

\begin{figure*}[!b]
    \centering
    \tikzset{every picture/.style={line width=0.75pt}} 

    \begin{tikzpicture}[x=0.75pt,y=0.75pt,yscale=-1,xscale=1]
    
    \draw [color={rgb, 255:red, 65; green, 117; blue, 5 }  ,draw opacity=1 ][line width=1.5]    (89.6,21.45) .. controls (91.08,23.35) and (90.88,25.05) .. (89.01,26.56) .. controls (87.12,27.99) and (86.89,29.64) .. (88.32,31.52) .. controls (89.84,33.23) and (89.77,34.85) .. (88.1,36.38) .. controls (86.73,38.43) and (87.13,40.05) .. (89.28,41.23) .. controls (91.52,41.86) and (92.41,43.28) .. (91.95,45.49) .. controls (91.84,47.91) and (92.96,49.12) .. (95.31,49.12) .. controls (97.72,49.09) and (98.98,50.21) .. (99.07,52.49) -- (101.33,54.33) ;
    \draw   (86.1,17.89) .. controls (86.1,15.93) and (87.69,14.33) .. (89.66,14.33) .. controls (91.62,14.33) and (93.22,15.93) .. (93.22,17.89) .. controls (93.22,19.86) and (91.62,21.45) .. (89.66,21.45) .. controls (87.69,21.45) and (86.1,19.86) .. (86.1,17.89) -- cycle ;
    \draw    (89.6,21.45) -- (68.6,49.15) ;
    \draw    (89.6,21.45) -- (109.6,49.15) ;
    \draw [color={rgb, 255:red, 65; green, 117; blue, 5 }  ,draw opacity=1 ][line width=1.5]    (148.6,20.92) .. controls (149.54,23.18) and (148.91,24.77) .. (146.71,25.68) .. controls (144.46,26.5) and (143.68,28.02) .. (144.37,30.23) .. controls (144.98,32.44) and (144.1,33.85) .. (141.72,34.45) .. controls (139.43,34.84) and (138.52,36.18) .. (138.99,38.46) .. controls (139.55,40.7) and (138.75,42.15) .. (136.59,42.81) .. controls (134.63,44.42) and (134.69,46.06) .. (136.77,47.73) .. controls (138.92,47.98) and (139.94,49.25) .. (139.84,51.53) .. controls (140.15,53.94) and (141.46,54.96) .. (143.77,54.58) -- (147.33,56.89) ;
    \draw   (145.69,18) .. controls (145.69,16.4) and (146.99,15.09) .. (148.6,15.09) .. controls (150.21,15.09) and (151.51,16.4) .. (151.51,18) .. controls (151.51,19.61) and (150.21,20.92) .. (148.6,20.92) .. controls (146.99,20.92) and (145.69,19.61) .. (145.69,18) -- cycle ;
    \draw    (148.6,20.92) -- (127.6,48.62) ;
    \draw    (148.6,20.92) -- (168.6,48.62) ;
    \draw [color={rgb, 255:red, 65; green, 117; blue, 5 }  ,draw opacity=1 ][line width=1.5]    (148.6,20.92) .. controls (150.84,22.35) and (151.41,24.06) .. (150.31,26.06) .. controls (149.17,27.95) and (149.65,29.46) .. (151.75,30.59) .. controls (153.81,31.62) and (154.28,33.21) .. (153.16,35.36) .. controls (152.03,37.53) and (152.46,39.18) .. (154.45,40.32) .. controls (156.42,41.55) and (156.73,43.15) .. (155.36,45.1) .. controls (153.82,46.79) and (153.79,48.43) .. (155.28,50.03) .. controls (155.39,52.69) and (154.18,53.71) .. (151.66,53.08) .. controls (149.21,52.43) and (147.87,53.36) .. (147.64,55.88) -- (147.33,56.33) ;
    \draw [color={rgb, 255:red, 65; green, 117; blue, 5 }  ,draw opacity=1 ][line width=1.5]    (201.6,21.92) .. controls (203.08,23.82) and (202.88,25.52) .. (201.01,27.02) .. controls (199.12,28.45) and (198.89,30.11) .. (200.32,31.98) .. controls (201.84,33.69) and (201.77,35.31) .. (200.1,36.84) .. controls (198.73,38.9) and (199.13,40.52) .. (201.28,41.69) .. controls (203.52,42.32) and (204.41,43.74) .. (203.95,45.95) .. controls (203.84,48.38) and (204.96,49.59) .. (207.31,49.59) .. controls (209.72,49.56) and (210.98,50.68) .. (211.07,52.95) -- (213.33,54.8) ;
    \draw   (198.1,18.36) .. controls (198.1,16.39) and (199.69,14.8) .. (201.66,14.8) .. controls (203.62,14.8) and (205.22,16.39) .. (205.22,18.36) .. controls (205.22,20.32) and (203.62,21.92) .. (201.66,21.92) .. controls (199.69,21.92) and (198.1,20.32) .. (198.1,18.36) -- cycle ;
    \draw    (201.6,21.92) -- (180.6,49.62) ;
    \draw    (201.6,21.92) -- (221.6,49.62) ;
    \draw   (241.1,39.36) .. controls (241.1,37.39) and (242.69,35.8) .. (244.66,35.8) .. controls (246.62,35.8) and (248.22,37.39) .. (248.22,39.36) .. controls (248.22,41.32) and (246.62,42.92) .. (244.66,42.92) .. controls (242.69,42.92) and (241.1,41.32) .. (241.1,39.36) -- cycle ;
    \draw    (244.6,42.92) -- (223.6,70.62) ;
    \draw    (244.6,42.92) -- (264.6,70.62) ;
    \draw [color={rgb, 255:red, 20; green, 99; blue, 192 }  ,draw opacity=1 ]   (213.33,54.8) .. controls (245.51,1.69) and (249.16,11.77) .. (244.99,34.06) ;
    \draw [shift={(244.66,35.8)}, rotate = 281.27] [color={rgb, 255:red, 20; green, 99; blue, 192 }  ,draw opacity=1 ][line width=0.75]    (10.93,-3.29) .. controls (6.95,-1.4) and (3.31,-0.3) .. (0,0) .. controls (3.31,0.3) and (6.95,1.4) .. (10.93,3.29)   ;
    \draw [color={rgb, 255:red, 65; green, 117; blue, 5 }  ,draw opacity=1 ][line width=1.5]    (298.6,18.92) .. controls (300.08,20.82) and (299.88,22.52) .. (298.01,24.02) .. controls (296.12,25.45) and (295.89,27.11) .. (297.32,28.98) .. controls (298.84,30.69) and (298.77,32.31) .. (297.1,33.84) .. controls (295.73,35.9) and (296.13,37.52) .. (298.28,38.69) .. controls (300.52,39.32) and (301.41,40.74) .. (300.95,42.95) .. controls (300.84,45.38) and (301.96,46.59) .. (304.31,46.59) .. controls (306.72,46.56) and (307.98,47.68) .. (308.07,49.95) -- (310.33,51.8) ;
    \draw   (295.1,15.36) .. controls (295.1,13.39) and (296.69,11.8) .. (298.66,11.8) .. controls (300.62,11.8) and (302.22,13.39) .. (302.22,15.36) .. controls (302.22,17.32) and (300.62,18.92) .. (298.66,18.92) .. controls (296.69,18.92) and (295.1,17.32) .. (295.1,15.36) -- cycle ;
    \draw    (298.6,18.92) -- (277.6,46.62) ;
    \draw    (298.6,18.92) -- (318.6,46.62) ;
    \draw   (338.1,36.36) .. controls (338.1,34.39) and (339.69,32.8) .. (341.66,32.8) .. controls (343.62,32.8) and (345.22,34.39) .. (345.22,36.36) .. controls (345.22,38.32) and (343.62,39.92) .. (341.66,39.92) .. controls (339.69,39.92) and (338.1,38.32) .. (338.1,36.36) -- cycle ;
    \draw    (341.6,39.92) -- (320.6,67.62) ;
    \draw    (341.6,39.92) -- (361.6,67.62) ;
    \draw [color={rgb, 255:red, 20; green, 99; blue, 192 }  ,draw opacity=1 ]   (341.66,32.8) .. controls (330.73,5.01) and (319.17,34.77) .. (311.19,50.18) ;
    \draw [shift={(310.33,51.8)}, rotate = 298.39] [color={rgb, 255:red, 20; green, 99; blue, 192 }  ,draw opacity=1 ][line width=0.75]    (10.93,-3.29) .. controls (6.95,-1.4) and (3.31,-0.3) .. (0,0) .. controls (3.31,0.3) and (6.95,1.4) .. (10.93,3.29)   ;
    \draw [color={rgb, 255:red, 65; green, 117; blue, 5 }  ,draw opacity=1 ][line width=1.5]    (440.6,70.92) .. controls (442.25,72.7) and (442.28,74.42) .. (440.69,76.08) .. controls (439.5,78.31) and (440.31,79.47) .. (443.12,79.56) .. controls (444.15,77.51) and (445.78,76.98) .. (447.99,77.96) .. controls (449.9,79.19) and (451.52,78.92) .. (452.86,77.13) .. controls (454.81,75.61) and (456.42,75.86) .. (457.71,77.89) .. controls (458.65,80.07) and (460.22,80.82) .. (462.41,80.15) -- (465,82) ;
    \draw   (392.1,27.36) .. controls (392.1,25.39) and (393.69,23.8) .. (395.66,23.8) .. controls (397.62,23.8) and (399.22,25.39) .. (399.22,27.36) .. controls (399.22,29.32) and (397.62,30.92) .. (395.66,30.92) .. controls (393.69,30.92) and (392.1,29.32) .. (392.1,27.36) -- cycle ;
    \draw    (395.6,30.92) -- (374.6,58.62) ;
    \draw    (395.6,30.92) -- (415.6,58.62) ;
    \draw   (435.1,48.36) .. controls (435.1,46.39) and (436.69,44.8) .. (438.66,44.8) .. controls (440.62,44.8) and (442.22,46.39) .. (442.22,48.36) .. controls (442.22,50.32) and (440.62,51.92) .. (438.66,51.92) .. controls (436.69,51.92) and (435.1,50.32) .. (435.1,48.36) -- cycle ;
    \draw    (438.6,51.92) -- (411,78) ;
    \draw    (438.6,51.92) -- (473,77) ;
    \draw [color={rgb, 255:red, 20; green, 99; blue, 192 }  ,draw opacity=1 ]   (399.22,27.36) .. controls (422.88,14.6) and (433.98,55.98) .. (439.81,69.24) ;
    \draw [shift={(440.6,70.92)}, rotate = 242.84] [color={rgb, 255:red, 20; green, 99; blue, 192 }  ,draw opacity=1 ][line width=0.75]    (10.93,-3.29) .. controls (6.95,-1.4) and (3.31,-0.3) .. (0,0) .. controls (3.31,0.3) and (6.95,1.4) .. (10.93,3.29)   ;
    \draw [color={rgb, 255:red, 65; green, 117; blue, 5 }  ,draw opacity=1 ][line width=1.5]    (527.6,66.92) .. controls (529.25,68.7) and (529.28,70.42) .. (527.69,72.08) .. controls (526.5,74.31) and (527.31,75.47) .. (530.12,75.56) .. controls (531.15,73.51) and (532.78,72.98) .. (534.99,73.96) .. controls (536.9,75.19) and (538.52,74.92) .. (539.86,73.13) .. controls (541.81,71.61) and (543.42,71.86) .. (544.71,73.89) .. controls (545.65,76.07) and (547.22,76.82) .. (549.41,76.15) -- (552,78) ;
    \draw   (485.1,19.36) .. controls (485.1,17.39) and (486.69,15.8) .. (488.66,15.8) .. controls (490.62,15.8) and (492.22,17.39) .. (492.22,19.36) .. controls (492.22,21.32) and (490.62,22.92) .. (488.66,22.92) .. controls (486.69,22.92) and (485.1,21.32) .. (485.1,19.36) -- cycle ;
    \draw    (488.6,22.92) -- (467.6,50.62) ;
    \draw    (488.6,22.92) -- (508.6,50.62) ;
    \draw   (528.1,40.36) .. controls (528.1,38.39) and (529.69,36.8) .. (531.66,36.8) .. controls (533.62,36.8) and (535.22,38.39) .. (535.22,40.36) .. controls (535.22,42.32) and (533.62,43.92) .. (531.66,43.92) .. controls (529.69,43.92) and (528.1,42.32) .. (528.1,40.36) -- cycle ;
    \draw    (531.6,43.92) -- (504,70) ;
    \draw    (531.6,43.92) -- (566,69) ;
    \draw [color={rgb, 255:red, 20; green, 99; blue, 192 }  ,draw opacity=1 ]   (492.22,19.36) .. controls (503,16.29) and (522.33,56.21) .. (526.77,65.25) ;
    \draw [shift={(527.6,66.92)}, rotate = 241.25] [color={rgb, 255:red, 20; green, 99; blue, 192 }  ,draw opacity=1 ][line width=0.75]    (10.93,-3.29) .. controls (6.95,-1.4) and (3.31,-0.3) .. (0,0) .. controls (3.31,0.3) and (6.95,1.4) .. (10.93,3.29)   ;
    \draw   (586.1,16.36) .. controls (586.1,14.39) and (587.69,12.8) .. (589.66,12.8) .. controls (591.62,12.8) and (593.22,14.39) .. (593.22,16.36) .. controls (593.22,18.32) and (591.62,19.92) .. (589.66,19.92) .. controls (587.69,19.92) and (586.1,18.32) .. (586.1,16.36) -- cycle ;
    \draw    (589.6,19.92) -- (568.6,47.62) ;
    \draw    (589.6,19.92) -- (609.6,47.62) ;
    \draw [color={rgb, 255:red, 20; green, 99; blue, 192 }  ,draw opacity=1 ]   (552,78) .. controls (538.42,43.08) and (565.3,26.98) .. (584.36,17.24) ;
    \draw [shift={(586.1,16.36)}, rotate = 153.22] [color={rgb, 255:red, 20; green, 99; blue, 192 }  ,draw opacity=1 ][line width=0.75]    (10.93,-3.29) .. controls (6.95,-1.4) and (3.31,-0.3) .. (0,0) .. controls (3.31,0.3) and (6.95,1.4) .. (10.93,3.29)   ;

    \draw (9.5,36) node[minirond] (A1) {};
    \node[above=0.1em of A1] {$\indva$};
    \draw (52,36) node[minirond] (A2) {};
    \node[above=0.1em of A2] {$\indvb$};
    \path[->] (A1) edge [color={rgb, 255:red, 65; green, 117; blue, 5 } ,draw opacity=1] node[yshift=8] {} (A2);

    \draw (95.67,6.93) node [anchor=north west][inner sep=0.75pt]    {$\indva$};
    \draw (153.67,8.4) node [anchor=north west][inner sep=0.75pt]    {$\indva$};
    \draw (207.67,7.4) node [anchor=north west][inner sep=0.75pt]    {$\indva$};
    \draw (250.67,28.4) node [anchor=north west][inner sep=0.75pt]    {$\indvo$};
    \draw (304.67,4.4) node [anchor=north west][inner sep=0.75pt]    {$\indva$};
    \draw (347.67,25.4) node [anchor=north west][inner sep=0.75pt]    {$\indvo$};
    \draw (382.67,13.4) node [anchor=north west][inner sep=0.75pt]    {$\indvo$};
    \draw (445.67,37.4) node [anchor=north west][inner sep=0.75pt]    {$\indva$};
    \draw (480,5.4) node [anchor=north west][inner sep=0.75pt]    {$\indvo$};
    \draw (535,29.4) node [anchor=north west][inner sep=0.75pt]    {$\indva$};
    \draw (580,-2.6) node [anchor=north west][inner sep=0.75pt]    {$\indvoo$};
    \end{tikzpicture}
    \vspace{-0.5em}
    \begin{tikzpicture}

        \draw (-8,0) node[minirond] (A1) {};
        \node[above=0.1em of A1] {$\indva$};
        \node[below=0.1em of A1,blue] {$\concept{D}_{\stateq, \stateq'}^{\automatonA}$};
        \draw (-6.85,0) node[minirond] (A2) {};
        \node[above=0.1em of A2] {$\indvb$};
        \path[->] (A1) edge [blue] node[yshift=8] {$\role{d}^{\automatonA}_{\stateq, \stateq'}$} (A2);

        \draw (-5.8,0) node[minirond] (B1) {};
        \path[->, >=stealth] (B1) edge [blue, loop left] node[yshift=11,xshift=28] {$\role{i}^{\automatonA}_{\stateq, \stateq'_\deadend}$} ();
        \node[right=0.1em of B1] {$\indva$};
        \node[below=0.1em of B1,blue] {$\concept{I}_{\stateq, \stateq'}^{\automatonA}$};

        \draw (-4.3,0) node[minirond] (C1) {};
        \path[->, >=stealth] (C1) edge [blue, loop right] node[yshift=10,xshift=-18] {$\role{rt}^{\automatonA}_{\stateq, \stateq'}$} ();
        \node[left=0.1em of C1] {$\indva$};
        \node[below=0.1em of C1,blue] {$\concept{RT}_{\stateq, \stateq'}^{\automatonA}$ \; \;};

        \draw (-3.3,0) node[minirond] (D1) {};
        \node[above=0.1em of D1] {$\indva$};
        \draw (-1.5,0) node[minirond] (D2) {};
        \node[above=0.1em of D2] {$\indvo$};
        \path[->] (D1) edge [blue] node[yshift=8] {$\role{io}^{\automatonA, \indvo}_{\stateq, \stateq'}$} (D2);
        \node[below=0.1em of D1,blue] {$\; \; \concept{IO}_{\stateq, \stateq'}^{\automatonA, \indvo}$};

        \draw (-0.8,0) node[minirond] (E1) {};
        \node[above=0.1em of E1] {$\indva$};
        \draw (1.2,0) node[minirond] (E2) {};
        \node[above=0.1em of E2] {$\indvo$};
        \path[->] (E2) edge [blue] node[yshift=8] {$\role{oi}^{\automatonA, \indvo}_{\stateq, \stateq'}$} (E1);
        \node[below=0.1em of E1,blue] {$\concept{OI}_{\stateq, \stateq'}^{\automatonA, \indvo}$};

        \draw (2.15,0) node[minirond] (F1) {};
        \node[above=0.1em of F1] {$\indvo$};
        \draw (3.9,0) node[minirond] (F2) {};
        \node[above=0.1em of F2] {$\indva$};
        \path[->] (F1) edge [blue] node[yshift=8] {$\role{i}^{\automatonA, \indvo}_{\stateq,  \stateq'_\deadend}$} (F2);

        \node[below=0.1em of F2,blue] {$\concept{I}_{\stateq, \stateq'}^{\automatonA, \indvo}$};

        \draw (4.75,0) node[minirond] (G1) {};
        \node[above=0.1em of G1] {$\indvo$};
        \draw (6.35,0) node[minirond] (G2) {};
        \node[above=0.1em of G2] {$\indvoo$};
        \path[->] (G1) edge [blue] node[yshift=8] {$\role{by}^{\automatonA, \indvo, \indvoo}_{\stateq,  \stateq'}$} (G2);

        \draw (7.15,0) node[minirond] (G3) {};
        \node[right=0.1em of G3] {$\indva$};
        \node[below=0.1em of G3, blue] {$\;\ \concept{By}^{\automatonA, \indvo, \indvoo}_{\stateq,  \stateq'}$};
    \end{tikzpicture}
\caption{Basic paths and their corresponding decorations in quasi-forests, given in order of their introduction in Def.~\ref{def:basic-paths-in-quasi-forests}, \ref{def:A-concepts}, and \ref{def:A-roles}.}\label{fig:basic-paths-in-quasi-forests}
\end{figure*}
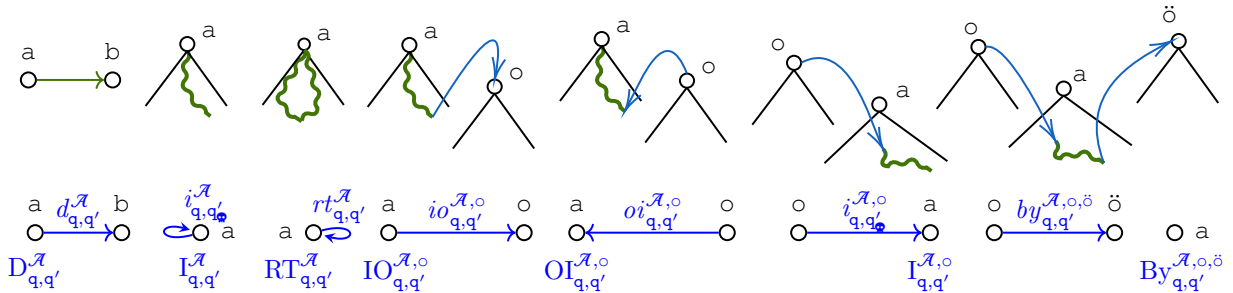

The main goal of this section is to provide a characterisation of how paths in quasi-forests look like and how they can be decomposed into interesting pieces. 
A path $\pathrho$ in $\interI$ is \emph{nameless} if it does not contain any \emph{named} elements (\ie no element of $\pathrho$ has the form $\indva^{\interI}$ for some $\indva \in \Ilang$).  
Similarly, $\pathrho$ is called an \emph{$\indva$-subtree path} if all the members of $\pathrho$ are descendants of~$\indva^{\interI}$. 
In quasi-forests, nameless paths are precisely subtree paths.

\begin{definition}\label{def:basic-paths-in-quasi-forests}
  Let $\interI$ be an $(\IlangA, \IlangT)$-quasi-forest, $\indva, \indvb \in (\IlangA \cup \IlangT)$, $\indvo, \indvoo \in \IlangT$, and $\pathrho$ be a path in~$\interI$. We call $\pathrho$:
  \begin{itemize}\itemsep0em\itemindent=0cm
    \item \vocab{$(\indva, \indvb)$-direct} if $\pathrho = \indva^{\interI}{\cdot}\indvb^{\interI}$.
    \item \vocab{$\indva$-inner} if $\pathrho = \indva^{\interI}{\cdot}\bar{\pathrho}$ for some $\indva$-subtree path $\bar{\pathrho}$.
    \item \vocab{$\indva$-roundtrip} if $\pathrho = \indva^{\interI}{\cdot}\bar{\pathrho}{\cdot}\indva^{\interI}$ for some $\indva$-subtree path $\bar{\pathrho}$.
    \item \vocab{$(\indva, \indvo)$-inout} if $\pathrho = \bar{\pathrho}{\cdot}\indvo^{\interI}$ for some $\indva$-inner path $\bar{\pathrho}$.
    \item \vocab{$(\indva, \indvo)$-outin} if the reverse of $\pathrho$ is $(\indva, \indvo)$-inout.
    \item \vocab{$(\indva, \indvo)$-inner} if $\pathrho = \indvo^{\interI}{\cdot}\bar{\pathrho}$ for some $\indva$-subtree path $\bar{\pathrho}$.
    \item \vocab{$(\indva, \indvo, \indvoo)$-bypass} if $\pathrho = \indvo^{\interI}{\cdot}\bar{\pathrho}{\cdot}\indvoo^{\interI}$ for an $\indva$-subtree path $\bar{\pathrho}$.
  \end{itemize}
  If $\pathrho$ falls into one of the above seven categories, we call $\pathrho$~\vocab{basic}. 
  Paths $\pathrho \deff \bar{\pathrho}{\cdot}\domelemd$ for a nameless $\bar{\pathrho}$ and a root $\domelemd$ are called \vocab{outer}.
  Finally, $\pathrho$ is \vocab{decomposable} whenever there exists a growing sequence of indices $i_1 < i_2 < \ldots < i_k$ with $i_1 {=} 1$ and $i_k {=} |\pathrho|$ such that for all $j < k$ the path $\pathrho_{i_j} \ldots \pathrho_{i_{(j{+}1)}}$ is basic.\myqed
\end{definition}

By a careful analysis of Definitions \ref{def:quasi-forest}--\ref{def:basic-paths-in-quasi-forests} we can show:
\begin{observation}\label{obs:basic-paths-ind-base}
  Take $\interI$, $\indva$, $\indvb$, $\pathrho$ as in \Cref{def:basic-paths-in-quasi-forests}. If $\pathrho$ has the form $\indva^{\interI}{\cdot}\bar{\pathrho}$ or $\indva^{\interI}{\cdot}\bar{\pathrho}{\cdot}\indvb^{\interI}$ for a nameless $\bar{\pathrho}$ then $\pathrho$ is basic.\myqed
\end{observation}

We can now employ Observation~\ref{obs:basic-paths-ind-base} and invoke the induction over the total number of named elements in $\pathrho$ to establish:
\begin{lemma}\label{lemma:basic-decomposition-of-paths}
  If $\interI$ is a quasi-forest then every path starting from a named element is decomposable.\myqed
\end{lemma}

Basic paths are expressible in $\ZOIQ$ by means of regular expressions with tests. 
Their construction involves nominal tests $\{ \indva \}?$, ``descendant of $\indva$'' tests $\exists{(\role{child}^-)^+}.\{ \indva \}?$, and role names $\role{edge}, \role{child}$. For instance, all $(\indva, \indvo, \indvoo)$-bypasses match:
\[  \{ \indvo \}? \ \role{edge}\ \left( [\exists{(\role{child}^-)^+}.\{ \indva \}]?\ \role{edge} \right)^+\  \{ \indvoo \}?
\]
\begin{lemma}\label{lemma:describing-basic-paths-in-ZOIQ}
Let $\tau$ be a  ``category'' of paths from Def.~\ref{def:basic-paths-in-quasi-forests} (including nameless and outer paths). 
There is an NFA $\automatonA_{\tau}$ (using testing involving only individual names mentioned in~$\tau$) that realises for all quasi-forests $\interI$ exactly the paths of the category~$\tau$. Moreover, for any regular language $\languageL$ (given as an NFA $\automatonB$ or a reg-exp $\regexpR$) there exists an NFA $(\automatonB \bowtie \automatonA_{\tau})$ (resp. $(\regexpR \bowtie \automatonA_{\tau})$) of size polynomial in $|\automatonB|$ (resp.  $|\regexpR|$) that realises precisely the paths in $\interI$ of category $\tau$ realising $\automatonB$ (resp. $\regexpR$).~\myqed
\end{lemma}


\section{Automata Decorations}\label{sec:automata-decorations}

\paragraph*{Overview.}
We use Lem.~\ref{lemma:basic-decomposition-of-paths} to devise a method of decorating the clearing of $\interI$ with extra information about the basic~paths realising an NFA $\automatonA$.
This helps to decide the satisfaction of concepts $\exists{\automatonA}.\top$ in quasi-forests in a modular way,  independently for the clearing of a quasi-forest and its~subtrees. Note:
\begin{fact}
  Let $\domelemd \in (\exists{\automatonA.\top})^{\interI}$ for an NFA $\automatonA$ and an $(\IlangA, \IlangT)$-quasi-forest $\interI$. There is an $\automatonA$-path $\pathrho$ starting from $\domelemd$ s.t.: (a) $\domelemd$ is a root, (b) $\pathrho$ is nameless, or (c) $\pathrho = \bar{\pathrho}{\cdot}\hat{\pathrho}$ for some~outer~$\bar{\pathrho}$.~\myqed
\end{fact}

Given an NFA $\automatonA$ with the state set $\statesQ$, by the \vocab{$\automatonA$ reachability concepts $\lang{C}_{\leadsto}(\automatonA)$} we mean the set $\textstyle\{ \concept{Reach}^{\automatonA}_{\stateq, \stateq'} \mid \stateq, \stateq' \in \statesQ \}$.
Ideally, $\textstyle\concept{Reach}^{\automatonA}_{\stateq, \stateq'}$ should label precisely the roots of quasi-forests  satisfying $\exists{\automatonA_{\stateq,\stateq'}}.\top$.
With \Cref{lemma:the-main-property-of-A-decorated-quasi-forests} we explain the desired ``modularity'' condition.
It simply says that if $\textstyle\concept{Reach}^{\automatonA}_{\stateq, \stateq'}$-concepts are interpreted as desired, then the verification of whether an element $\domelemd$ satisfies $\exists{\automatonA_{\stateq,\stateq'}}.\top$ boils to (i) testing whether $\domelemd$ is labelled with $\textstyle\concept{Reach}^{\automatonA}_{\stateq, \stateq'}$ if $\domelemd$ is a root, or (ii) testing existence of a certain path which fully contained in the subtree of $\domelemd$, possibly except the last element which can also be the root of $\domelemd$ or a nominal.
We rely on the NFAs $\automatonA_{\textrm{nmls}}$ and $\automatonA_{\textrm{outr}}$ from \Cref{lemma:describing-basic-paths-in-ZOIQ} that detect nameless and outer paths. 

\begin{lemma}\label{lemma:the-main-property-of-A-decorated-quasi-forests}
  Let $\automatonA$ be an NFA with the set of states $\statesQ$, and let $\interI$ be an $(\IlangA, \IlangT)$-quasi-forest that interprets all $\textstyle\concept{Reach}^{\automatonA}_{\stateq, \stateq'}$-concepts from $\lang{C}_{\leadsto}(\automatonA)$ equally to $\concept{Root} \dland \exists{\automatonA_{\stateq,\stateq'}}.\top$.
  Then for all states $\stateq, \stateq' \in \statesQ$, the concept $\exists{\automatonA_{\stateq, \stateq'}}.\top$ is interpreted in $\interI$ equally to the~union~of concepts:\\
  $\bullet$ $\concept{Root} \dland \concept{Reach}_{\stateq, \stateq'}$,\; \;
  $\bullet$ $\neg\concept{Root} \dland \exists{\left( \automatonA_{\stateq, \stateq'} \bowtie \automatonA_{\textrm{nmls}} \right)}.\top$, and\\
  $\bullet$ $\neg\concept{Root} \dland \bigdlor_{\hat{\stateq} \in \statesQ} \exists{\left( \automatonA_{\stateq, \hat{\stateq}} \bowtie \automatonA_{\textrm{outr}}\right)}.\left(\concept{Root} \dland \concept{Reach}_{\hat{\stateq}, \stateq'}\right)$.\myqed
\end{lemma}

\noindent Let \vocab{$\textstyle\mathrm{rel}(\IlangT, \automatonA_{\stateq, \stateq'})$} denote the above concept union, called the \vocab{$\textstyle\IlangT$-relativisation} of $\exists{\automatonA_{\stateq, \stateq'}}.\top$.
We say that~$\textstyle\domelemd \in \DeltaI$ \vocab{virtually satisfies} $\exists{\automatonA_{\stateq, \stateq'}}.\top$ whenever $\domelemd$ satisfies~$\mathrm{rel}(\IlangT, \automatonA_{\stateq, \stateq'})$.
Thus, \Cref{lemma:the-main-property-of-A-decorated-quasi-forests} tells us that under some extra assumptions the notion of satisfaction and virtual satisfaction coincide.

\paragraph*{Decorations.}
In what follows, we define $\ZOIQ$-concepts and roles, dubbed \emph{automata decorations}, intended to guarantee the desired interpretation of the reachability concepts.

\begin{definition}\label{def:A-concepts}
  Given an NFA $\automatonA$ with the state-set $\statesQ$, the set of \vocab{$(\IlangT, \automatonA)$-concepts $\concepts(\IlangT,\automatonA)$} is composed of the following concept names (for all states $\stateq, \stateq' \in \statesQ$ and names $\indvo, \indvoo \in \IlangT$):
  \[  \concept{D}_{\stateq, \stateq'}^{\automatonA}, \;
      \concept{I}_{\stateq, \stateq'}^{\automatonA}, \;
      \concept{RT}_{\stateq, \stateq'}^{\automatonA}, \;
      \concept{IO}_{\stateq, \stateq'}^{\automatonA, \indvo}, \; 
      \concept{OI}_{\stateq, \stateq'}^{\automatonA, \indvo}, \; 
      \concept{I}_{\stateq, \stateq'}^{\automatonA, \indvo}, \;
      \concept{By}^{\automatonA, \indvo, \indvoo}_{\stateq,  \stateq'}.
  \] 
  A quasi-forest $\interI$ \vocab{properly interprets} $\concepts(\IlangT,\automatonA)$ if  $\conceptC^{\interI} = \{ \indva^{\interI} \mid \indva \in \Ilang, \mathrm{cond}(\conceptC, \indva) \}$ for concepts and conditions as below.
  \vspace{-0.6em}
  \begin{table}[H]
    \centering
  \rowcolors{2}{White}{LightBlue!30}
  \renewcommand{\arraystretch}{1.2}
  \begin{tabular}{l|l}\toprule
  conc. $\conceptC$  & $\mathrm{cond}(\conceptC, \indva)\colon$ ``There is a path $\pathrho \models  \automatonA_{\stateq, \stateq'}$ s.t.  \\\midrule
   $\concept{D}_{\stateq, \stateq'}^{\automatonA}$ & $\pathrho$ is $(\indva,\indvb)$-direct for some $\indvb^{\interI}$''. \\
   $\concept{I}_{\stateq, \stateq'}^{\automatonA}$ &  $\pathrho$ is $\indva$-inner''. \\
   $\concept{RT}_{\stateq, \stateq'}^{\automatonA}$ &  $\pathrho$ is an $\indva$-roundtrip''. \\
   $\concept{IO}_{\stateq, \stateq'}^{\automatonA, \indvo}$ &  $\pathrho$ is an $(\indva, \indvo)$-inout''. \\
   $\concept{OI}_{\stateq, \stateq'}^{\automatonA, \indvo}$ &  $\pathrho$ is an $(\indva, \indvo)$-outin''.\\
   $\concept{I}_{\stateq, \stateq'}^{\automatonA, \indvo}$ &  $\pathrho$ is $(\indva, \indvo)$-inner''.\\
   $\concept{By}^{\automatonA, \indvo, \indvoo}_{\stateq,  \stateq'}$&  $\pathrho$ is an $(\indva, \indvo, \indvoo)$-bypass''.\\\bottomrule
  \end{tabular}
  \renewcommand{\arraystretch}{1}
  \end{table}
  \vspace{-1em}
  \noindent The size of $\concepts(\IlangT,\automatonA)$ is clearly polynomial in~$|\automatonA|{\cdot}|\IlangT|$.~\myqed
\end{definition}

The concepts from $\concepts(\IlangT,\automatonA)$ ``bookkeep'' the information about relevant basic paths realising NFA starting from a given~root. 
Throughout the paper, we employ a \emph{fresh} individual name $\ghost$ (the \emph{ghost variable}), to stress that the constructed concepts are independent from $\IlangA$.
With $\conceptC[\ghost/\indva]$ we denote the result of substituting $\indva$ for all occurrences of~$\ghost$~in~$\conceptC$.
Relying on NFAs from Lemma~\ref{lemma:describing-basic-paths-in-ZOIQ}, we construct $\ZOIQ$-concepts that describe the intended behaviour of $(\IlangT, \automatonA)$-concepts. 
For instance, $\concept{RT}_{\stateq, \stateq'}^{\automatonA}$ is defined as $\textstyle\exists{\left( \automatonA_{\stateq,\stateq'} \bowtie \automatonA_{\ghost\text{-roundtrip}} \right)}.\top$.

\begin{lemma}\label{lemma:intended-behaviour-of-A-concepts}
  For all $\conceptC$ from $\concepts(\IlangT,\automatonA)$, there exists a $\ZOIQ$-concept $\mathrm{desc}(\conceptC)$  (of size polynomial w.r.t. $|\automatonA|{\cdot}|\IlangT|$) that uses only individual names from $\IlangT {\cup} \{ \ghost \}$ with the property that ``for all $(\IlangA, \IlangT)$-quasi-forests $\interI$ and~$\indva \in \Ilang$ we have:
  {\color{white}..}$\mathrm{cond}(\conceptC, \indva)$ is satisfied in $\interI$ iff $\indva^{\interI}$ is in $\left( \mathrm{desc}(\conceptC)[\ghost/\indva] \right)^{\interI}$.''\myqed
\end{lemma}

 We next invoke \Cref{lemma:intended-behaviour-of-A-concepts} to rephrase the notion of proper interpretation of $\concepts(\IlangT,\automatonA)$ in the language of satisfaction of $\ZOIQ$-concepts by the clearings of quasi-forests. We have:

\begin{fact}\label{corr:intended-behaviour-of-A-concepts}
  Let $\mathrm{comdsc}(\IlangT, \automatonA) \deff \textstyle\bigdland_{\conceptC \in \concepts(\IlangT,\automatonA)}(\conceptC {\leftrightarrow} \mathrm{desc}(\conceptC))$ An $(\IlangA, \IlangT)$-quasi-forest~$\interI$ properly interprets $\concepts(\IlangT,\automatonA)$ iff $\indva^{\interI} \in \mathrm{comdsc}(\IlangT, \automatonA)\left[ \ghost/\indva \right]^{\interI}$  for all $\indva \in (\IlangA \cup \IlangT)$.\myqed
\end{fact}

The information provided by the proper interpretation of $(\IlangT, \automatonA)$-concepts is not sufficient to ensure the proper interpretation of the reachability concepts by the clearings of quasi-forests. 
The main reason is that the concepts from $(\IlangT, \automatonA)$ concern only about the basic paths. As some automata constraints cannot be satisfied by basic paths, \eg $\exists{\left(\{\indvo\}? \role{edge} \{\indvoo\}? \role{edge} \{  \indvvo\}?\right)}.\top$, more work needs to be done.
To be able to compose basic paths into bigger pieces, another ``layer of decoration'' is needed: this time with fresh roles representing the endpoints of basic~paths from~\Cref{def:basic-paths-in-quasi-forests}, as summarised by \Cref{fig:basic-paths-in-quasi-forests}.
For instance, whenever there is an $(\indva, \indvo)$-inout $\pathrho$ realising $\automatonA_{\stateq,\stateq'}$ in a quasi-forest $\interI$, the roots $\indva^{\interI}$ and $\indvo^{\interI}$ are going to be linked by the role $\role{io}^{\automatonA, \indvo}_{\stateq, \stateq'}$ (and similarly for other categories of basic paths that start and end in roots).
Such roles can be seen as ``aggregated paths''. The ``aggregated paths'' will be unfolded afterwards into real paths, and the satisfaction of automata constraints by them will be verified by means of the \emph{guided automata} (introduced in \Cref{A-guided-automaton}).
However, there exist cases of basic paths where the last element is unnamed, more precisely $(\indva, \indvo)$-inner and $\indva$-inner paths. In their cases we treat $\indva^{\interI}$ as the ``virtual end'' of $\pathrho$ (and hence we link $\indva^{\interI}$ and $\indvo^{\interI}$). 
To ensure that such an ``aggregated path'' can no longer be extended, we decorate the corresponding role with the dead-end symbol~$\deadend$. When constructing the guided automaton, the roles carrying $\deadend$ will lead to states with no outgoing transitions (dead ends).

\begin{definition}\label{def:A-roles}
  Given a PSA $\automatonA$ with the state-set $\statesQ$, the set of \vocab{$(\IlangT, \automatonA)$-roles $\lang{R}(\IlangT,\automatonA)$} is composed of the following role names (for all states $\stateq, \stateq' \in \statesQ$ and names $\indvo, \indvoo \in \IlangT$):
  \[ 
  \role{d}^{\automatonA}_{\stateq, \stateq'}, \;
  \role{i}^{\automatonA}_{\stateq, \stateq'_\deadend}, \;
  \role{rt}^{\automatonA}_{\stateq, \stateq'}, \;
  \role{io}^{\automatonA, \indvo}_{\stateq, \stateq'}, \; 
  \role{oi}^{\automatonA, \indvo}_{\stateq, \stateq'}, \; 
  \role{i}^{\automatonA, \indvo}_{\stateq,  \stateq'_\deadend}, \;
  \role{by}^{\automatonA, \indvo, \indvoo}_{\stateq,  \stateq'}.
  \]
  An $(\IlangA, \IlangT)$-quasi-forest $\interI$ \vocab{properly interprets} $\lang{R}(\IlangT,\automatonA)$ if its roles consist of pairs $\mathrm{p}$ of roots of $\interI$ as indicated below.
  \vspace{-0.6em}
  \begin{table}[H]
    \centering
  \rowcolors{2}{White}{LightBlue!30}
  \renewcommand{\arraystretch}{1.3}
  \begin{tabular}{lll}\toprule
   Role name $\roler$ & Pair $\mathrm{p}$ & Condition \\\midrule
   $\role{d}^{\automatonA}_{\stateq, \stateq'}$ & $(\indva^\interI,\indvb^\interI)$ & $\indva^{\interI}\indvb^{\interI}$ $\models \automatonA_{\stateq,\stateq'}$ \\
   $\role{i}^{\automatonA}_{\stateq, \stateq'_\deadend}$ & $(\indva^\interI,\indva^\interI)$ & $\indva^{\interI}$ is in $(\concept{I}_{\stateq, \stateq'}^{\automatonA})^{\interI}$ \\
   $\role{rt}^{\automatonA}_{\stateq, \stateq'}$  & $(\indva^\interI,\indva^\interI)$ & $\indva^{\interI}$ is in $(\concept{RT}_{\stateq, \stateq'}^{\automatonA})^{\interI}$ \\
   $\role{io}^{\automatonA, \indvo}_{\stateq, \stateq'}$ & $(\indva^\interI,\indvo^\interI)$ & $\indva^{\interI}$ is in $(\concept{IO}_{\stateq, \stateq'}^{\automatonA, \indvo})^{\interI}$ \\
   $\role{oi}^{\automatonA, \indvo}_{\stateq, \stateq'}$ & $(\indvo^\interI,\indva^\interI)$  & $\indva^{\interI}$ is in $(\concept{OI}_{\stateq, \stateq'}^{\automatonA, \indvo})^{\interI}$ \\
   $\role{i}^{\automatonA, \indvo}_{\stateq,  \stateq'_\deadend}$ & $(\indvo^\interI,\indva^\interI)$ & $\indva^{\interI}$ is in $(\concept{I}_{\stateq, \stateq'}^{\automatonA, \indvo})^{\interI}$ \\
   $\role{by}^{\automatonA, \indvo, \indvoo}_{\stateq, \stateq'}$ & $(\indvo^\interI,\indvoo^\interI)$ & $(\concept{By}^{\automatonA, \indvo, \indvoo}_{\stateq,  \stateq'})^{\interI}$ is non-empty \\\bottomrule
  \end{tabular}
  \renewcommand{\arraystretch}{1}
\end{table}
\vspace{-0.6em}
\noindent The size of $\lang{R}(\IlangT,\automatonA)$ is polynomial w.r.t. $|\automatonA|{\cdot}|\IlangT|$.\myqed
\end{definition}

\noindent We stress that we interpreted role names from $\textstyle\lang{R}(\IlangT,\automatonA)$ based on the concepts from $\textstyle\lang{C}(\IlangT,\automatonA)$ rather than on the existence of certain paths realising $\automatonA$. 
This allows us to verify their proper interpretation, independently from the verification of the proper interpretation of  $\textstyle\lang{C}(\IlangT,\automatonA)$-concepts.
The roles from $\textstyle\lang{R}(\IlangT,\automatonA)$ can be seen as ``shortcuts'' aggregating fragments of runs~of~$\automatonA$.
To retrieve the runs, we invoke the $\automatonA$-guided automaton that operates solely on such~``shortcuts''. 

\begin{definition}\label{A-guided-automaton}
  For an NFA $\automatonA$ with the state-set $\statesQ$, we define the \vocab{$\automatonA$-guided NFA $\automatonB$}.
  The set of states $\statesQ'$ of $\automatonB$ consists of all $\stateq \in \statesQ$ and their fresh copies~$\stateq_\deadend$.
  The transitions in $\automatonB$ have the form $(\stateq, \roler, \stateq')$ for all $\stateq,\stateq' \in \statesQ'$, and all $\roler$ from $\lang{R}(\IlangT,\automatonA)$ labelled with the ordered pair $(\stateq, \stateq')$. By design, all states decorated with $\deadend$ have no outgoing transitions. \myqed
\end{definition}

The guided automaton $\automatonB$ is of size polynomial in~$|\automatonA|{\cdot}|\IlangT|$, and it traverses only the clearings of quasi-forests. The following Lemma~\ref{lemma:main-property-of-A-guided-automata} reveals the desired property of $\automatonB$.

\begin{lemma}\label{lemma:main-property-of-A-guided-automata}
  Let $\automatonA$ be an NFA with an $\automatonA$-guided NFA $\automatonB$, and $\interI$ be an $(\IlangA, \IlangT)$-quasi-forest that properly interprets $\concepts(\IlangT,\automatonA)$ and $\lang{R}(\IlangT,\automatonA)$. For all states $\stateq, \stateq'$ of~$\automatonA$:
  $
    (\concept{Root} \dland \exists{\automatonA_{\stateq, \stateq'}.\top})^{\interI} =
    (\concept{Root} \dland (\exists{\automatonB_{\stateq, \stateq'}.\top} \dlor \exists{\automatonB_{\stateq, \stateq'_\deadend}.\top}))^{\interI}
  $.~\myqed
\end{lemma}

Its proof relies on either (a) shortening a path $\pathrho \models \automatonA_{\stateq,  \stateq'}$ to its subsequence composed of all named elements, or (b) replacing any two consecutive elements in~$\pathrho \models \automatonB_{\stateq,  \stateq'}$ with the corresponding paths guaranteed by the roles from~Def.~\ref{def:A-roles}.

\begin{example}\label{example:guided-automata}
  Let $\pathrho \deff \indva^{\interI}{\cdot}\indvo^{\interI}{\cdot}\pathrho_1{\cdot}\indvo^{\interI}\pathrho_2{\cdot}\indvoo^{\interI}{\cdot}\pathrho_3 \models \automatonA_{\stateq_1, \stateq_5}$ be a path in~a quasi-forest $\interI$ (depicted below), and $\automatonA$ be an NFA with states indicated by decorated $\stateq$. Suppose (i) $\indva^{\interI}{\cdot}\indvo^{\interI}\models \automatonA_{\stateq_1, \stateq_2}$ is $(\indva,\indvo)$-direct, (ii) $\indvo^{\interI}{\cdot}\pathrho_1{\cdot}\indvo^{\interI} \models \automatonA_{\stateq_2, \stateq_3}$ is an $\indvo$-roundtrip,~(iii) $\indvo^{\interI}\pathrho_2{\cdot}\indvoo^{\interI} \models \automatonA_{\stateq_3,\stateq_4}$ is an $(\indvb, \indvo, \indvoo)$-bypass, and (iv) $\indvoo^{\interI}{\cdot}\pathrho_3 \models \automatonA_{\stateq_4, \stateq_5}$ is $(\indvc, \indvoo)$-inner. So,~$\indva^{\interI} \in (\exists{\automatonA_{\stateq_1, \stateq_5}}.\top)^{\interI}$.
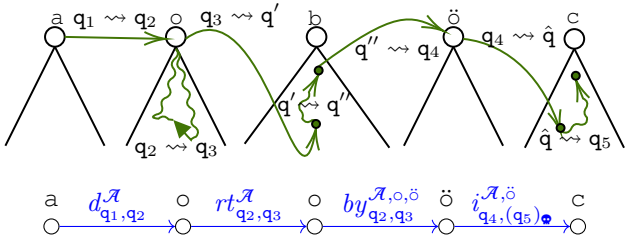
\vspace{-1.5em}

\begin{figure}[H]
    \centering
    \tikzset{every picture/.style={line width=0.75pt}} 

    \begin{tikzpicture}[x=0.75pt,y=0.75pt,yscale=-1,xscale=1]
    
    \draw   (28.33,15.06) .. controls (28.33,12.33) and (30.55,10.12) .. (33.27,10.12) .. controls (36,10.12) and (38.22,12.33) .. (38.22,15.06) .. controls (38.22,17.79) and (36,20) .. (33.27,20) .. controls (30.55,20) and (28.33,17.79) .. (28.33,15.06) -- cycle ;
    \draw    (33.27,20) -- (58.33,70) ;
    \draw    (33.27,20) -- (8.33,70) ;
    
    \draw   (89.33,15.06) .. controls (89.33,12.33) and (91.55,10.12) .. (94.27,10.12) .. controls (97,10.12) and (99.22,12.33) .. (99.22,15.06) .. controls (99.22,17.79) and (97,20) .. (94.27,20) .. controls (91.55,20) and (89.33,17.79) .. (89.33,15.06) -- cycle ;
    \draw    (94.27,20) -- (119.33,70) ;
    \draw    (94.27,20) -- (69.33,70) ;
    
    \draw   (160.33,15.06) .. controls (160.33,12.33) and (162.55,10.12) .. (165.27,10.12) .. controls (168,10.12) and (170.22,12.33) .. (170.22,15.06) .. controls (170.22,17.79) and (168,20) .. (165.27,20) .. controls (162.55,20) and (160.33,17.79) .. (160.33,15.06) -- cycle ;
    \draw    (165.27,20) -- (202,71) ;
    \draw    (165.27,20) -- (129,69) ;
    \draw   (229.33,15.06) .. controls (229.33,12.33) and (231.55,10.12) .. (234.27,10.12) .. controls (237,10.12) and (239.22,12.33) .. (239.22,15.06) .. controls (239.22,17.79) and (237,20) .. (234.27,20) .. controls (231.55,20) and (229.33,17.79) .. (229.33,15.06) -- cycle ;
    \draw    (234.27,20) -- (259.33,70) ;
    \draw    (234.27,20) -- (209.33,70) ;
    
    \draw   (290.33,16.06) .. controls (290.33,13.33) and (292.55,11.12) .. (295.27,11.12) .. controls (298,11.12) and (300.22,13.33) .. (300.22,16.06) .. controls (300.22,18.79) and (298,21) .. (295.27,21) .. controls (292.55,21) and (290.33,18.79) .. (290.33,16.06) -- cycle ;
    \draw    (295.27,21) -- (320.33,71) ;
    \draw    (295.27,21) -- (270.33,71) ;
    
    \draw [color={rgb, 255:red, 65; green, 117; blue, 5 }  ,draw opacity=1 ][line width=0.75]    (38.22,15.06) -- (87.33,16.02) ;
    \draw [shift={(89.33,16.06)}, rotate = 181.12] [color={rgb, 255:red, 65; green, 117; blue, 5 }  ,draw opacity=1 ][line width=0.75]    (10.93,-3.29) .. controls (6.95,-1.4) and (3.31,-0.3) .. (0,0) .. controls (3.31,0.3) and (6.95,1.4) .. (10.93,3.29)   ;
    \draw [color={rgb, 255:red, 65; green, 117; blue, 5 }  ,draw opacity=1 ][line width=0.75]    (94.27,21) .. controls (95.8,22.86) and (95.66,24.53) .. (93.84,26.02) .. controls (91.96,27.72) and (91.74,29.4) .. (93.19,31.07) .. controls (94.54,33.06) and (94.25,34.7) .. (92.3,35.98) .. controls (90.27,37.32) and (89.85,38.99) .. (91.04,40.98) .. controls (92.17,42.95) and (91.64,44.51) .. (89.44,45.68) .. controls (87.29,46.43) and (86.59,47.92) .. (87.32,50.15) .. controls (87.89,52.4) and (86.98,53.75) .. (84.57,54.18) .. controls (81.89,55.33) and (82.22,56.05) .. (85.56,56.34) .. controls (87.39,54.76) and (89.07,54.87) .. (90.6,56.68) -- (93.01,56.97) ;
    \draw [color={rgb, 255:red, 65; green, 117; blue, 5 }  ,draw opacity=1 ][line width=0.75]    (94.27,21.56) .. controls (96.36,22.89) and (96.82,24.61) .. (95.66,26.74) .. controls (94.49,28.8) and (94.91,30.4) .. (96.94,31.55) .. controls (98.95,32.64) and (99.39,34.36) .. (98.28,36.71) .. controls (97.03,38.5) and (97.43,40.07) .. (99.49,41.41) .. controls (101.52,42.66) and (101.92,44.26) .. (100.69,46.23) .. controls (99.42,48.08) and (99.8,49.67) .. (101.83,50.98) .. controls (103.82,52.17) and (104.19,53.8) .. (102.94,55.87) .. controls (101.62,57.72) and (101.92,59.25) .. (103.85,60.46) .. controls (105.76,61.88) and (105.98,63.53) .. (104.49,65.41) .. controls (102.96,65.96) and (102.12,66.51) .. (101.97,67.06) -- (100.11,65.13) -- (95,59.16) ;
    \draw [shift={(93.01,56.97)}, rotate = 46.72] [fill={rgb, 255:red, 65; green, 117; blue, 5 }  ,fill opacity=1 ][line width=0.08]  [draw opacity=0] (8.93,-4.29) -- (0,0) -- (8.93,4.29) -- cycle    ;
    \draw [color={rgb, 255:red, 65; green, 117; blue, 5 }  ,draw opacity=1 ]   (99.22,16.06) .. controls (138.82,-13.64) and (155.66,114.4) .. (164.73,62.62) ;
    \draw [shift={(165,61)}, rotate = 99.13] [color={rgb, 255:red, 65; green, 117; blue, 5 }  ,draw opacity=1 ][line width=0.75]    (10.93,-3.29) .. controls (6.95,-1.4) and (3.31,-0.3) .. (0,0) .. controls (3.31,0.3) and (6.95,1.4) .. (10.93,3.29)   ;
    \draw [color={rgb, 255:red, 65; green, 117; blue, 5 }  ,draw opacity=1 ][line width=0.75]    (165.34,35.88) -- (161.38,42.94) .. controls (162.04,45.15) and (161.23,46.6) .. (158.96,47.28) .. controls (156.69,48.11) and (156,49.64) .. (156.91,51.89) .. controls (158.37,53.26) and (158.74,54.72) .. (158.02,56.27) .. controls (159.41,58.38) and (161.07,58.65) .. (163,57.08) -- (165,57) ;
    \draw [shift={(166.33,34)}, rotate = 117.51] [color={rgb, 255:red, 65; green, 117; blue, 5 }  ,draw opacity=1 ][line width=0.75]    (10.93,-3.29) .. controls (6.95,-1.4) and (3.31,-0.3) .. (0,0) .. controls (3.31,0.3) and (6.95,1.4) .. (10.93,3.29)   ;
    \draw [color={rgb, 255:red, 65; green, 117; blue, 5 }  ,draw opacity=1 ]   (166.33,30) .. controls (204.93,1.05) and (212.8,3.76) .. (227.68,11.23) ;
    \draw [shift={(229.33,12.06)}, rotate = 206.73] [color={rgb, 255:red, 65; green, 117; blue, 5 }  ,draw opacity=1 ][line width=0.75]    (10.93,-3.29) .. controls (6.95,-1.4) and (3.31,-0.3) .. (0,0) .. controls (3.31,0.3) and (6.95,1.4) .. (10.93,3.29)   ;
    \draw [color={rgb, 255:red, 65; green, 117; blue, 5 }  ,draw opacity=1 ]   (239.22,16.06) .. controls (263.38,18.6) and (280.15,37.68) .. (287.77,61.18) ;
    \draw [shift={(288.33,63)}, rotate = 253.23] [color={rgb, 255:red, 65; green, 117; blue, 5 }  ,draw opacity=1 ][line width=0.75]    (10.93,-3.29) .. controls (6.95,-1.4) and (3.31,-0.3) .. (0,0) .. controls (3.31,0.3) and (6.95,1.4) .. (10.93,3.29)   ;
    \draw [color={rgb, 255:red, 65; green, 117; blue, 5 }  ,draw opacity=1 ][line width=0.75]    (296.72,38.6) -- (299.67,46.15) .. controls (301.81,47.13) and (302.35,48.67) .. (301.28,50.78) .. controls (299.97,52.61) and (300.09,54.26) .. (301.64,55.71) .. controls (301.94,58.33) and (300.82,59.55) .. (298.29,59.37) .. controls (296.19,58.53) and (294.67,59.18) .. (293.74,61.32) .. controls (292.64,63.43) and (291.02,63.94) .. (288.87,62.85) -- (288.33,63) ;
    \draw [shift={(296,36.67)}, rotate = 69.78] [color={rgb, 255:red, 65; green, 117; blue, 5 }  ,draw opacity=1 ][line width=0.75]    (10.93,-3.29) .. controls (6.95,-1.4) and (3.31,-0.3) .. (0,0) .. controls (3.31,0.3) and (6.95,1.4) .. (10.93,3.29)   ;
    \draw  [fill={rgb, 255:red, 65; green, 117; blue, 5 }  ,fill opacity=1 ] (163,59) .. controls (163,57.9) and (163.9,57) .. (165,57) .. controls (166.1,57) and (167,57.9) .. (167,59) .. controls (167,60.1) and (166.1,61) .. (165,61) .. controls (163.9,61) and (163,60.1) .. (163,59) -- cycle ;
    \draw  [fill={rgb, 255:red, 65; green, 117; blue, 5 }  ,fill opacity=1 ] (164.33,32) .. controls (164.33,30.9) and (165.23,30) .. (166.33,30) .. controls (167.44,30) and (168.33,30.9) .. (168.33,32) .. controls (168.33,33.1) and (167.44,34) .. (166.33,34) .. controls (165.23,34) and (164.33,33.1) .. (164.33,32) -- cycle ;
    \draw  [fill={rgb, 255:red, 65; green, 117; blue, 5 }  ,fill opacity=1 ] (286.33,61) .. controls (286.33,59.9) and (287.23,59) .. (288.33,59) .. controls (289.44,59) and (290.33,59.9) .. (290.33,61) .. controls (290.33,62.1) and (289.44,63) .. (288.33,63) .. controls (287.23,63) and (286.33,62.1) .. (286.33,61) -- cycle ;
    \draw  [fill={rgb, 255:red, 65; green, 117; blue, 5 }  ,fill opacity=1 ] (294,34.67) .. controls (294,33.56) and (294.9,32.67) .. (296,32.67) .. controls (297.1,32.67) and (298,33.56) .. (298,34.67) .. controls (298,35.77) and (297.1,36.67) .. (296,36.67) .. controls (294.9,36.67) and (294,35.77) .. (294,34.67) -- cycle ;
    
    \draw (41,2) node [anchor=north west][inner sep=0.75pt]    {\small{$\stateq_1 \leadsto \stateq_2$}};
    \draw (72,67) node [anchor=north west][inner sep=0.75pt] {\small{$\stateq_2 \leadsto \stateq_3$}};
    \draw (105,-3.5) node [anchor=north west][inner sep=0.75pt] {\small{$\stateq_3 \leadsto \stateq'$}};
    \draw (144,41.4) node [anchor=north west][inner sep=0.75pt]    {\small{$\stateq' {\leadsto}\ \stateq''$}};
    \draw (183,13) node [anchor=north west][inner sep=0.75pt]     {\small{$\stateq'' \leadsto \stateq_4$}};
    \draw (247,8) node [anchor=north west][inner sep=0.75pt]    {\small{$\stateq_4 \leadsto \hat{\stateq}$}};
    \draw (276,60) node [anchor=north west][inner sep=0.75pt] {\small{$\hat{\stateq} \leadsto \stateq_5$}};

    \draw (30,2) node [anchor=north west][inner sep=0.75pt]    {\small{$\indva$}};
    \draw (90,2) node [anchor=north west][inner sep=0.75pt]    {\small{$\indvo$}};
    \draw (160,0) node [anchor=north west][inner sep=0.75pt]    {\small{$\indvb$}};
    \draw (230,0) node [anchor=north west][inner sep=0.75pt]    {\small{$\indvoo$}};
    \draw (290,1.5) node [anchor=north west][inner sep=0.75pt]    {\small{$\indvc$}};
    \end{tikzpicture}
\end{figure}
\vspace{-4.5em}
\begin{figure}[H]
    \centering
    \begin{tikzpicture}
        \draw (0,0) node[minirond] (A1) {};
        \node[above=0.1em of A1] {$\indva$};
        \draw (1.75,0) node[minirond] (A2) {};
        \node[above=0.1em of A2] {$\indvo$};
        \path[->] (A1) edge [blue] node[yshift=8] {$\role{d}^{\automatonA}_{\stateq_1, \stateq_2}$} (A2);
        \draw (3.5,0) node[minirond] (A3) {};
        \node[above=0.1em of A3] {$\indvo$};
        \path[->] (A2) edge [blue] node[yshift=8] {$\role{rt}^{\automatonA}_{\stateq_2, \stateq_3}$} (A3);
        \draw (5.25,0) node[minirond] (A4) {};
        \node[above=0.1em of A4] {$\indvoo$};
        \path[->] (A3) edge [blue] node[yshift=8] {$\role{by}^{\automatonA, \indvo, \indvoo}_{\stateq_2, \stateq_3}$} (A4);
        \draw (7,0) node[minirond] (A5) {};
        \node[above=0.1em of A5] {$\indvc$};
        \path[->] (A4) edge [blue] node[yshift=8] {$\role{i}^{\automatonA, \indvoo}_{\stateq_4,  (\stateq_5)_\deadend}$} (A5);
    \end{tikzpicture}
\end{figure}
\vspace{-1em}
\noindent Consider now the $\automatonA$-guided $\automatonB$, and observe that $\indva^{\interI}$ is in $(\exists{\automatonB_{\stateq_1, (\stateq_5)_\deadend}}.\top)^{\interI}$, as witnessed by the path $\indva^{\interI}{\cdot}\indvo^{\interI}{\cdot}\indvo^{\interI}{\cdot}\indvoo^{\interI}{\cdot}\indvc^{\interI}$ and the word $\role{d}^{\automatonA}_{\stateq_1, \stateq_2}{\cdot}\role{rt}^{\automatonA}_{\stateq_2, \stateq_3}{\cdot}\role{by}^{\automatonA, \indvo, \indvoo}_{\stateq_2, \stateq_3}{\cdot}\role{i}^{\automatonA, \indvoo}_{\stateq_4,  (\stateq_5)_\deadend}$.~\myqed
\end{example}

With the proviso that a quasi-forest $\interI$ properly interprets both $\concepts(\IlangT,\automatonA)$ and $\lang{R}(\IlangT,\automatonA)$, \Cref{lemma:main-property-of-A-guided-automata} guarantees the desired interpretation of reachability concepts $\concept{Reach}^{\automatonA}_{\stateq, \stateq'}$.

\begin{definition}\label{def:A-decoration}
  Let $\interI, \automatonA, \automatonB$ be as in \Cref{lemma:main-property-of-A-guided-automata}, and $\statesQ$ be the state-set of $\automatonA$. The set of \vocab{$\automatonA$-reachability-concepts $\lang{C}_{\leadsto}(\automatonA)$} is $\textstyle\{ \concept{Reach}^{\automatonA}_{\stateq, \stateq'} \mid \stateq, \stateq' \in \statesQ \}$.
  $\interI$ \vocab{properly interprets} $\lang{C}_{\leadsto}(\automatonA)$ if
  \vspace{-0.5em}
  \begin{center}$\textstyle(\concept{Reach}_{\stateq, \stateq'}^{\automatonA})^{\interI} =  (\concept{Root} \dland (\exists{\automatonB_{\stateq, \stateq'}.\top} \dlor \exists{\automatonB_{\stateq, \stateq'_\deadend}.\top}))^{\interI}$\end{center}
  \vspace{-0.5em}
  for all $\stateq, \stateq' \in \statesQ$.
  We call $\interI$ \vocab{virtually $\textstyle(\IlangT,\automatonA)$-decorated} if it properly interprets $\lang{R}(\IlangT,\automatonA)$, and $\textstyle\lang{C}_{\leadsto}(\automatonA)$, and \vocab{$(\IlangT,\automatonA)$-decorated} if it additionally properly interprets~$\textstyle\concepts(\IlangT,\automatonA)$.~\myqed
\end{definition}

Relying on \Cref{lemma:the-main-property-of-A-decorated-quasi-forests} and \Cref{lemma:main-property-of-A-guided-automata} we can show:

\begin{lemma}\label{lemma:the-second-to-last-lemma-on-automata-decorations}
  Let $\automatonA$ be an NFA, and let $\interI$ be an $(\IlangT,\automatonA)$-decorated $(\IlangA, \IlangT)$-quasi-forest. For all elements $\domelemd$ in~$\interI$, if $\domelemd$ virtually satisfies $\exists{\automatonA_{\stateq, \stateq'}}.\top$ then $\domelemd$ satisfies~$\exists{\automatonA_{\stateq, \stateq'}}.\top$.
\end{lemma}

We conclude by showing an algorithmic result concerning (virtually) $(\IlangT,\automatonA)$-decorated quasi-forests. It exploits the fact that the regular path queries over databases can be evaluated in $\PTime$~\cite[Lemma~3.1]{MendelzonW95}.

\begin{lemma}\label{lemma:testing-A-decoration-in-PTime}
  Let $\interI$ be a finite interpretation with $\concept{Root}^{\interI} = \DeltaI$, and $\automatonA$ be an NFA.
  We can then verify in time polynomial w.r.t. $(|\automatonA| {\cdot} |\interI|)$ whether $\interI$ is virtually $(\IlangT,\automatonA)$-decorated.\myqed
\end{lemma}

\noindent \Cref{lemma:testing-A-decoration-in-PTime} tells us that if $\interI$ is the clearing of some $(\IlangA, \IlangT)$-quasi-forest $\interJ$ that properly interprets $\concepts(\IlangT,\automatonA)$-concepts, then we can verify if $\interJ$~is~$(\IlangT,\automatonA)$-decorated~in~$\PTime$.

\section{Counting Decorations}\label{sec:counting-decorations}
We want to ``relativise'' number restrictions in the presence of nominals, so that in the suitably decorated quasi-forest models, the satisfaction of concepts of the form $({\geq}{n}\; \roler).\top$ by the clearing can be decided solely based on the decoration of the clearing. 
Observe that for a given a root $\domelemd$ of a quasi-forest $\interI$ and a role name $\roler$, the set of $\roler$-successors of $\domelemd$ can be divided into three groups: (a) the clearing, (b) the children of~$\domelemd$, and (c) the descendants of roots (but only in case $\domelemd$ is a nominal root).  
To relativise counting, we decorate each element of the clearing with the total number of their $\roler$-successors in categories (a) and (b), as well as the information, for each nominal~$\indvo$, on (c) how many of their descendants are $\roler$-successors of~$\indvo$.

\begin{definition}\label{def:IT-r-n-counting-decoration}
  Fix $\roler \in \Rlang$, $n \in \N$, and a finite $\IlangT \subseteq \Ilang$.~The set~of \vocab{$(\IlangT, {\geq}{n}\ \roler)$-counting-concepts $\concepts_{\#}(\IlangT,{\geq}{n}\ \roler)$} consists of concept names $\concept{Clrng}_{\mathfrak{t}}^{\roler}$, $\concept{Chld}^{\roler}_{\mathfrak{t}}$, $\concept{Des}_{\mathfrak{t}}^{\roler, \indvo}$ for $\indvo \in \IlangT$ and thresholds~$\mathfrak{t}$ of the form ``${=}m$'' for $0 \leq m \leq n$ or~``${\geq}n{+}1$''.
  All~integers appearing in thresholds are encoded in binary.
  An $(\IlangA, \IlangT)$-quasi-forest $\interI$ is \vocab{$({\geq}{n}\ \roler)$-semi-decorated} if all roots $\domelemd$ have unique thresholds $\homot_{\text{cl}}^{\domelemd}$, $\homot_{\text{ch}}^{\domelemd}$, 
  and $\homot_{\indvo}^{\domelemd}$ for all $\indvo \in \IlangT$, for which $\domelemd$ satisfies the concept  
  \begin{align*}
    \vspace{-0.5em}
    \concept{Clrng}_{\homot_{\text{cl}}^{\domelemd}}^{\roler} \dland \concept{Chld}_{\homot_{\text{ch}}^{\domelemd}}^{\roler} \dland \textstyle\bigdland_{\indvo \in \IlangT} \concept{Des}_{\homot_{\indvo}^{\domelemd} }^{\roler, \indvo}.
    \vspace{-0.5em}
  \end{align*}
  \vspace{-0.03em}
  The above concepts are called the \vocab{$(\IlangT,{\geq}{n}\ \roler)$-descriptions of~$\domelemd$}. Note that their size is polynomial w.r.t. $|\IlangT|{\cdot}\log_2(n)$.\myqed
\end{definition}

\noindent Similarly to the previous section, we next introduce the notion of proper satisfaction.
Intuitions regarding \Cref{def:IT-r-n-counting-decoration} are provided in \Cref{example:counting-decorations}.
We encourage the reader to read it.

\begin{definition}\label{def:IT-r-n-properly-interprets}
  An $(\IlangA, \IlangT)$-quasi-forest $\interI$ \vocab{properly interprets} $\concepts_{\#}(\IlangT,{\geq}{n}\ \roler)$ if $\interI$ is $({\geq}{n}\ \roler)$-semi-decorated and for all roots $\indva^{\interI}$ of $\interI$ and $\indvo \in \IlangT$, the root $\indva^{\interI}$ belongs to:\\
$\bullet$ $(\concept{Clrng}^{\roler}_{\mathfrak{t}})^{\interI}$ iff $|\{ \domelemd \colon \domelemd \in \concept{Root}^{\interI}, (\indva^{\interI}, \domelemd) \in \roler^{\interI} \}| \ \mathfrak{t}$,\\
$\bullet$ $(\concept{Chld}^{\roler}_{\mathfrak{t}})^{\interI}$ iff $|\{ \domelemd \colon (\indva^{\interI}, \domelemd) \in \roler^{\interI}, (\indva^{\interI}, \domelemd) \in \role{child}^{\interI} \}| \ \mathfrak{t}$,\\
$\bullet$ $(\concept{Des}^{\roler, \indvo}_{\mathfrak{t}})^{\interI}$ ifif $|\{\domelemd \colon (\indvo^{\interI}, \domelemd) \in \roler^{\interI}, (\indva^{\interI}, \domelemd) \in (\role{child}^+)^\interI  \}| \ \mathfrak{t}$.\\
In this case we also say that $\interI$ is \vocab{$({\geq}{n}\ \roler)$-decorated}.\myqed
\end{definition}

Based on the above definitions, we can easily express the intended behaviour of $\concepts_{\#}(\IlangT,{\geq}{n}\ \roler)$-concepts in $\ZOIQ$, using $({=}n\ \roler).\conceptC$ as an abbreviation of $({\geq}n\ \roler).\conceptC \dland \neg({\geq}n{+}1\ \roler).\conceptC$.
In~the most difficult case, we describe $(\concept{Des}^{\roler, \indvo}_{{=}42})^{\interI}$ with: \[\{ \ghost \} \dland \exists{\role{edge}^*}.\left( \{ \indvo \} \dland ({=}42 \; \roler).[\exists{(\role{child}^-)^+}.\{\ghost\}] \right).\]
\begin{lemma}\label{lemma:r-n-counting-decoration}
  For every $(\IlangT,{\geq}{n}\ \roler)$-description $\conceptC$ there is a $\ZOIQ$-concept $\mathrm{desc}(\conceptC)$ (of size polynomial w.r.t. $|\conceptC|$) that uses only individual names from $\IlangT \cup \{ \ghost \}$ for a fresh $\ghost$, s.t. for all $(\IlangT,{\geq}{n}\ \roler)$-semi-decorated $(\IlangA, \IlangT)$-quasi-forests  $\interI$:
  $\interI$ is $({\geq}{n}\ \roler)$-decorated iff for every root $\indva^{\interI}$ of $\interI$ and its $(\IlangT,{\geq}{n}\ \roler)$-description $\conceptC$ we have $\indva^{\interI} \in (\mathrm{desc}(\conceptC)[\ghost/\indva])^{\interI}$.\myqed
\end{lemma}

\begin{example}\label{example:counting-decorations}
  Let us consider an $(\{ \indva, \indvb \}, \{ \indvo \})$-quasi-forest $\interI$ sketched below and a role name $\roler$ depicted as a green arrow.
  As suggested by the drawing, (i) $\indva^{\interI}$ has no $\roler$-successors among the clearing of $\interI$ and precisely one $\roler$-successor among its children, (ii) $\indvo^{\interI}$ has two $\roler$-successors among the clearing of $\interI$, it has precisely two $\roler$-successor among its children, one $\roler$-successor that is a descendant of $\indva^{\interI}$, and three $\roler$-successors that are descendants of $\indvb^{\interI}$, and (iii) $\indvb^{\interI}$ has precisely one $\roler$-successor inside the clearing of $\interI$ and no other $\roler$-successors.

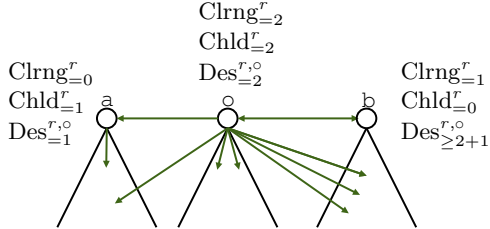
\begin{figure}[H]
    \centering
    \tikzset{every picture/.style={line width=0.75pt}} 
    \vspace{-0.5em}
    \begin{tikzpicture}[x=0.75pt,y=0.75pt,yscale=-1,xscale=1]
    
    \draw   (28.33,15.06) .. controls (28.33,12.33) and (30.55,10.12) .. (33.27,10.12) .. controls (36,10.12) and (38.22,12.33) .. (38.22,15.06) .. controls (38.22,17.79) and (36,20) .. (33.27,20) .. controls (30.55,20) and (28.33,17.79) .. (28.33,15.06) -- cycle ;
    \draw    (33.27,20) -- (58.33,70) ;
    \draw    (33.27,20) -- (8.33,70) ;
    
    \draw   (89.33,15.06) .. controls (89.33,12.33) and (91.55,10.12) .. (94.27,10.12) .. controls (97,10.12) and (99.22,12.33) .. (99.22,15.06) .. controls (99.22,17.79) and (97,20) .. (94.27,20) .. controls (91.55,20) and (89.33,17.79) .. (89.33,15.06) -- cycle ;
    \draw    (94.27,20) -- (119.33,70) ;
    \draw    (94.27,20) -- (69.33,70) ;
    
    \draw [color={rgb, 255:red, 62; green, 101; blue, 23 }  ,draw opacity=1 ]   (41.22,15.06) -- (89.33,15.06) ;
    \draw [shift={(38.22,15.06)}, rotate = 0] [fill={rgb, 255:red, 62; green, 101; blue, 23 }  ,fill opacity=1 ][line width=0.08]  [draw opacity=0] (3.57,-1.72) -- (0,0) -- (3.57,1.72) -- cycle    ;
    \draw [color={rgb, 255:red, 62; green, 101; blue, 23 }  ,draw opacity=1 ]   (102.22,15.06) -- (157.33,15.06) ;
    \draw [shift={(160.33,15.06)}, rotate = 180] [fill={rgb, 255:red, 62; green, 101; blue, 23 }  ,fill opacity=1 ][line width=0.08]  [draw opacity=0] (3.57,-1.72) -- (0,0) -- (3.57,1.72) -- cycle    ;
    \draw [shift={(99.22,15.06)}, rotate = 0] [fill={rgb, 255:red, 62; green, 101; blue, 23 }  ,fill opacity=1 ][line width=0.08]  [draw opacity=0] (3.57,-1.72) -- (0,0) -- (3.57,1.72) -- cycle    ;
    \draw [color={rgb, 255:red, 62; green, 101; blue, 23 }  ,draw opacity=1 ]   (89.73,38.09) -- (94.27,20) ;
    \draw [shift={(89,41)}, rotate = 284.1] [fill={rgb, 255:red, 62; green, 101; blue, 23 }  ,fill opacity=1 ][line width=0.08]  [draw opacity=0] (3.57,-1.72) -- (0,0) -- (3.57,1.72) -- cycle    ;
    \draw [color={rgb, 255:red, 62; green, 101; blue, 23 }  ,draw opacity=1 ]   (99.21,38.11) -- (94.27,20) ;
    \draw [shift={(100,41)}, rotate = 254.75] [fill={rgb, 255:red, 62; green, 101; blue, 23 }  ,fill opacity=1 ][line width=0.08]  [draw opacity=0] (3.57,-1.72) -- (0,0) -- (3.57,1.72) -- cycle    ;
    \draw [color={rgb, 255:red, 62; green, 101; blue, 23 }  ,draw opacity=1 ]   (33.04,37) -- (33.27,20) ;
    \draw [shift={(33,40)}, rotate = 270.79] [fill={rgb, 255:red, 62; green, 101; blue, 23 }  ,fill opacity=1 ][line width=0.08]  [draw opacity=0] (3.57,-1.72) -- (0,0) -- (3.57,1.72) -- cycle    ;
    \draw   (159.33,15.06) .. controls (159.33,12.33) and (161.55,10.12) .. (164.27,10.12) .. controls (167,10.12) and (169.22,12.33) .. (169.22,15.06) .. controls (169.22,17.79) and (167,20) .. (164.27,20) .. controls (161.55,20) and (159.33,17.79) .. (159.33,15.06) -- cycle ;
    \draw    (164.27,20) -- (189.33,70) ;
    \draw    (164.27,20) -- (139.33,70) ;
    
    \draw [color={rgb, 255:red, 62; green, 101; blue, 23 }  ,draw opacity=1 ]   (152.88,61.27) -- (94.27,20) ;
    \draw [shift={(155.33,63)}, rotate = 215.15] [fill={rgb, 255:red, 62; green, 101; blue, 23 }  ,fill opacity=1 ][line width=0.08]  [draw opacity=0] (3.57,-1.72) -- (0,0) -- (3.57,1.72) -- cycle    ;
    \draw [color={rgb, 255:red, 62; green, 101; blue, 23 }  ,draw opacity=1 ]   (161.5,43.03) -- (94.27,20) ;
    \draw [shift={(164.33,44)}, rotate = 198.91] [fill={rgb, 255:red, 62; green, 101; blue, 23 }  ,fill opacity=1 ][line width=0.08]  [draw opacity=0] (3.57,-1.72) -- (0,0) -- (3.57,1.72) -- cycle    ;
    \draw [color={rgb, 255:red, 62; green, 101; blue, 23 }  ,draw opacity=1 ]   (39.83,56.33) -- (94.27,20) ;
    \draw [shift={(37.33,58)}, rotate = 326.28] [fill={rgb, 255:red, 62; green, 101; blue, 23 }  ,fill opacity=1 ][line width=0.08]  [draw opacity=0] (3.57,-1.72) -- (0,0) -- (3.57,1.72) -- cycle    ;
    \draw [color={rgb, 255:red, 62; green, 101; blue, 23 }  ,draw opacity=1 ]   (161.5,43.03) -- (94.27,20) ;
    \draw [shift={(164.33,44)}, rotate = 198.91] [fill={rgb, 255:red, 62; green, 101; blue, 23 }  ,fill opacity=1 ][line width=0.08]  [draw opacity=0] (3.57,-1.72) -- (0,0) -- (3.57,1.72) -- cycle    ;
    \draw [color={rgb, 255:red, 62; green, 101; blue, 23 }  ,draw opacity=1 ]   (158.32,52.32) -- (94.27,20) ;
    \draw [shift={(161,53.67)}, rotate = 206.77] [fill={rgb, 255:red, 62; green, 101; blue, 23 }  ,fill opacity=1 ][line width=0.08]  [draw opacity=0] (3.57,-1.72) -- (0,0) -- (3.57,1.72) -- cycle    ; 
    
    \draw (29,2) node [anchor=north west][inner sep=0.75pt]    {\small{$\indva$}};
    \draw (90,2) node [anchor=north west][inner sep=0.75pt]    {\small{$\indvo$}};
    \draw (161,0) node [anchor=north west][inner sep=0.75pt]    {\small{$\indvb$}};

    \draw (-18,-15) node [anchor=north west][inner sep=0.75pt]    {\small{$\concept{Clrng}_{{=}0}^{\roler}$}};
    \draw (-18,0) node [anchor=north west][inner sep=0.75pt]    {\small{$\concept{Chld}_{{=}1}^{\roler}$}};
    \draw (-18,15) node [anchor=north west][inner sep=0.75pt]    {\small{$\concept{Des}_{{=}1}^{\roler, \indvo}$}};

    \draw (78,-15) node [anchor=north west][inner sep=0.75pt]    {\small{$\concept{Des}_{{=}2}^{\roler, \indvo}$}};
    \draw (78,-30) node [anchor=north west][inner sep=0.75pt]    {\small{$\concept{Chld}_{{=}2}^{\roler}$}};
    \draw (78,-45) node [anchor=north west][inner sep=0.75pt]    {\small{$\concept{Clrng}_{{=}2}^{\roler}$}};

    \draw (180,-15) node [anchor=north west][inner sep=0.75pt]    {\small{$\concept{Clrng}_{{=}1}^{\roler}$}};
    \draw (180,0) node [anchor=north west][inner sep=0.75pt]    {\small{$\concept{Chld}_{{=}0}^{\roler}$}};
    \draw (180,15) node [anchor=north west][inner sep=0.75pt]    {\small{$\concept{Des}_{{\geq}2{+}1}^{\roler, \indvo}$}};
    \end{tikzpicture}
    \vspace{-0.5em}
\end{figure}

  \noindent Suppose now that $\interI$ is $(\{ \indvo \},{\geq}{2}\ \roler)$-decorated. This implies:\\
    $\bullet\;\;\indva^{\interI} \in (\concept{Clrng}_{{=}0}^{\roler} \dland \concept{Chld}_{{=}1}^{\roler} \dland \concept{Des}_{{=}1}^{\roler, \indvo})^{\interI}$.\\
    $\bullet\;\;\indvo^{\interI} \in (\concept{Clrng}_{{=}2}^{\roler} \dland \concept{Chld}_{{=}2}^{\roler} \dland \concept{Des}_{{=}2}^{\roler, \indvo})^{\interI}$.\\
    $\bullet\;\;\indvb^{\interI} \in (\concept{Clrng}_{{=}1}^{\roler} \dland \concept{Chld}_{{=}0}^{\roler} \dland \concept{Des}_{{\geq}2{+}1}^{\roler, \indvo})^{\interI}$.\myqed
\end{example}

\Cref{lemma:r-n-counting-decoration}  rephrases the notion of proper satisfaction in the language of $\ZOIQ$-concepts.
Call a quasi-forest $\interI$ \vocab{virtually $(\IlangT,{\geq}{n}\ \roler)$-decorated} if $\interI$ is $(\IlangT,{\geq}{n}\ \roler)$-semi-decorated and properly interprets concepts of the form $\concept{Clrng}^{\roler}_{\mathfrak{t}}$. We have:
\begin{lemma}\label{lemma:verification-of-semi-decoration-in-PTime}
  For a finite interpretation $\interI$ we can test in polynomial time w.r.t. $|\interI|$ if $\interI$ is virtually $(\IlangT,{\geq}{n}\ \roler)$-decorated.~\myqed
\end{lemma} 

We use the numbers appearing in $(\IlangT,{\geq}{n}\ \roler)$-descriptions labelling the roots of $({\geq}{n}\ \roler)$-decorated quasi-forests to decide the satisfaction of $({\geq}{n}\ \roler).\top$.
We first describe it with an example. Suppose that we want to verify if a non-nominal root $\domelemd$ from $\interI$ satisfies $({\geq}{3}\ \roler)$ based on its labels $\concept{Clrng}^{\roler}_{{=}1}$ and $\concept{Chld}^{\roler}_{\geq 3{+}1}$. It suffices to check if $1 + (3+1)$ is at least~$3$. For the nominals roots~$\indvo$, we additionally take all the concepts $\concept{Des}^{\roler, \indvo}_{\mathfrak{t}}$~into~account, that are spread across the whole clearing.\\
An element $\domelemd$ from a finite $\interI$ \vocab{virtually satisfies}~$({\geq}{n}\ \roler).\top$, whenever $\domelemd$ satisfies $({\geq}{n}\ \roler).\top$ in every $({\geq}{n}\ \roler)$-decorated $(\IlangA, \IlangT)$-quasi-forest with the clearing equal to~$\interI$. We show:

\begin{lemma}\label{lemma:deciding-counting-GCIs-in-PTIme}
  For a finite virtually $(\IlangT,{\geq}{n}\ \roler)$-decorated interpretation $\interI$ with $\DeltaI = \concept{Root}^{\interI}$ and $\domelemd \in \concept{Root}^{\interI}$~we~can test in time polynomial in~$|\interI|$ if $\domelemd$ virtually satisfies~$({\geq}{n}\ \roler).\top$.\myqed
\end{lemma}

\section{Elegant Models and Their Summaries}\label{sec:summaries-of-quasi-forests}

In this section we benefit from various decorations introduced in the previous sections to design a succinct way of representing quasi-forest models of $\ZOIQ$-KBs, dubbed \emph{summaries}. 
From now on we will focus only on KBs in Scott's normal form (\ie with TBoxes in Scott's normal form) and on certain class of \emph{elegant} models introduced in \Cref{def:elegant-quasi-forest-models}.

\begin{definition}\label{def:elegant-quasi-forest-models}
  Let  $\kbK \deff (\aboxA, \tboxT)$ be a $\ZOIQ$-KB in Scott's normal form and let $\interI$ be its model.\\ 
  We call $\interI$ \vocab{elegant} if it satisfies all the conditions below:
  \begin{enumerate}[(i)]\itemsep0em
    \item $\interI$ is a canonical quasi-forest model of $\kbK$,
    \item $\interI$ is $(\indT, {\geq}{n}\ \roler)$-decorated for all number restrictions $({\geq}{n}\ \roler).\top$ from~$\tboxT$, 
    \item $\interI$ is $(\indT, \automatonA)$-decorated for all automata $\automatonA$ from~$\tboxT$,
    \item $\interI$ interprets all concept and role names that do not appear in $\kbK$, the set $\{ \concept{Root}, \role{edge}, \role{child}, \role{id} \}$, and in~mentioned~decorations, as the empty set.\myqed
  \end{enumerate}
\end{definition}

\noindent Invoking the normal form lemma [Appendix A], \Cref{lemma:quasi-forest-models-by-Ortiz-et-al}, as well as the definitions of proper interpretation, we obtain:

\begin{lemma}\label{lemma:elegant-models-for-scotts-normal-form}
  For every $\ZOIQ$-TBox $\tboxT$ we can compute in $\PTime$ a $\ZOIQ$-TBox $\tboxT'$ in Scott's normal form that possibly such that for every ABox $\aboxA$ we have:
  (i) $(\aboxA, \tboxT)$ is quasi-forest satisfiable iff $(\aboxA, \tboxT')$ has an elegant model, and 
  (ii) for every {\UCQ} $\queryq$ using only concepts and roles present in $\tboxT$ we have that $(\aboxA, \tboxT)$ has a canonical quasi-forest model violating~$\queryq$ iff $(\aboxA, \tboxT')$ has an elegant model violating $\queryq$.\myqed
  \end{lemma}

\noindent To-be-defined summaries are nothing more than ABoxes representing the full descriptions of clearings of elegant models.

\begin{definition}\label{def:complete-description-of-clearings}
  Let $\kbK \deff (\aboxA, \tboxT)$ be a $\ZOIQ$-KB in Scott's normal form.
  A \vocab{$\kbK$-summary $\aboxS$} is any $\subseteq$-minimal ABox satisfying, for all names $\indva, \indvb \in \indK$, all the conditions below.
  \begin{enumerate}[(I)]\itemsep0em
    \item $\aboxS$ contains either $\indva \approx \indvb$ or $\neg(\indva \approx \indvb)$.
    \item For all concept names $\conceptA$ appearing in $\kbK$ we have that $\aboxS$ contains either $\conceptA(\indva)$ or~$\neg\conceptA(\indva)$.
    
    \item For all NFA $\automatonA$ from $\kbK$, and all concept names $\conceptA$~from $\concepts(\indT,\automatonA) \cup \lang{C}_{\leadsto}(\automatonA)$, $\aboxS$ contains $\conceptA(\indva)$ or $\neg\conceptA(\indva)$.
    
    \item For all role names $\roler$ appearing in $\kbK$ we have that $\aboxS$ contains either $\roler(\indva, \indvb)$ or $\neg\roler(\indva, \indvb)$.
    
    \item For all NFA $\automatonA$ from $\kbK$, and all role names $\roler$ from $\lang{R}(\indT,\automatonA)$, the ABox $\aboxS$ contains $\roler(\indva, \indvb)$~or~$\neg\roler(\indva, \indvb)$.

    \item For all number restrictions $({\geq}{n}\ \roler).\top$ from $\kbK$, and for all $\indvo \in \indT$, there are thresholds $\homot_{\text{cl}}^{\indva}$, $\homot_{\text{ch}}^{\indva}$, 
    and $\homot_{\indvo}^{\indva}$ in $\{ {=}m, {\geq}n{+}1 \mid 0 \leq m \leq n\}$ for which $\aboxS$ contains 
    $\concept{Clrng}_{\homot_{\text{cl}}^{\indva}}^{\roler}(\indva)$, $\concept{Chld}_{\homot_{\text{ch}}^{\indva}}^{\roler}(\indva)$, and $\concept{Des}_{\homot_{\indvo}^{\indva} }^{\roler, \indvo}(\indva)$.

    \item $\concept{Root}(\indva) \in \abox{S}$ and $\role{edge}(\indva, \indvb) \in \abox{S}$.
  \end{enumerate}
    In the case when only a $\ZOIQ$-TBox $\tboxT$ is given, we define \vocab{$\tboxT$-ghost-summaries} as $(\{ \ghost \approx \ghost \}, \tboxT)$-summaries.\myqed
\end{definition}

Invoking the previously-established bounds on the number and sizes of concepts and roles occurring in decorations (\ie the ones stated at the end of Definitions \ref{def:A-concepts}, \ref{def:A-roles}, and \ref{def:IT-r-n-counting-decoration}),~we~infer:
\begin{lemma}\label{lemma:bounding-sizes-and-numbers-of-summaries}
  Let $\kbK$ and $\tboxT$ be, respectively, a $\ZOIQ$-KB and a $\ZOIQ$-TBox, both in Scott's normal form.
  We have that every $\kbK$-summary is of size polynomial in~$|\kbK|$. 
  Moreover, there are exponentially (in $|\tboxT|$) many $\tboxT$-ghost-summaries and each of them is of size polynomial w.r.t. $|\tboxT|$.\myqed
\end{lemma}

\noindent While a $\kbK$-summary can be easily extracted from the clearing of any elegant model of $\kbK$, the converse direction requires two extra assumptions, dubbed \emph{clearing-} and \emph{subtree-consistency}.


\section{Consistent Summaries}\label{sec:consistent-summaries-of-quasi-forests}

The notion of \emph{clearing-consistency} ensures that a given summary $\aboxS$ is a good candidate for the clearing of some elegant model of $\kbK$, namely (i) $\aboxS$ does not violate ``local'' constraints from $\kbK$, (ii) $\aboxS$ is virtually decorated for decorations involving number restrictions and automata, and that (iii) every element from~$\aboxS$ virtually satisfies all concepts involving automata or number restrictions from $\kbK$.
Below, $\interI_{\aboxS}$ denotes the minimal interpretation that corresponds to and satisfies the~ABox~$\aboxS$.

\begin{definition}\label{def:clearing-consistent}
  Let $\kbK \deff (\aboxA, \tboxT)$ be a $\ZOIQ$-KB in Scott's normal form, and let $\aboxS$ be a $\kbK$-summary.
  Let $\tboxTloc$ be the TBox composed of all GCIs from $\tboxT$ except for the ones concerning NFAs and number restrictions.
  We say that $\aboxS$ is \vocab{clearing-consistent} if $\interI_{\aboxS}$ satisfies $(\aboxA, \tboxTloc)$, and:\\
  $\bullet$ For all GCIs $\conceptA \equiv \exists{\automatonA_{\stateq,\stateq'}.\top}$ from $\tboxT$: (i) $\interI_{\aboxS}$ is virtually $(\ind{\tboxT},\automatonA)$-decorated, and (ii) for all $\indva \in \indK$, $\conceptA(\indva) \in \aboxS$ iff $\indva$ virtually satisfies $\exists{\automatonA_{\stateq,\stateq'}}.\top$ in $\interI_{\aboxS}$.\\
  $\bullet$ For all GCIs $\conceptA \equiv ({\geq}{n}\ \roler).\top$ from $\tboxT$: (i) $\interI_{\aboxS}$ is virtually $(\ind{\tboxT},{\geq}{n}\ \roler)$-decorated, and (ii) for all $\indva \in \indK$, we have that $\conceptA(\indva) \in \aboxS$ iff $\indva$ virtually satisfies $({\geq}{n}\ \roler).\top$ in $\interI_{\aboxS}$.
  \myqed
\end{definition}

Based on \Cref{lemma:testing-A-decoration-in-PTime} and \Cref{lemma:deciding-counting-GCIs-in-PTIme} we can conclude that deciding clearing consistency can be done in~$\PTime$.
\begin{lemma}\label{lemma:deciding-clearing-consistency-is-in-PTime}
  For $\kbK$ and $\aboxS$ as in Def.~\ref{def:clearing-consistent} we can decide in time polynomial w.r.t. $|\kbK|\cdot|\aboxS|$ whether $\aboxS$ is clearing consistent.\myqed
\end{lemma}

The second required notion is the \emph{subtree-consistency}. 
To decide whether a given summary $\aboxS$ extends to a model of $\kbK$, we need to check, for all  elements $\domelemd$ of $\aboxS$, the existence of a suitable forest satisfying (a relativised) $\kbK$ and fulfilling all the premises given by the decorations of $\domelemd$. This is achieved by crafting a suitable $\ZOIQ$-KB and testing whether it has a quasi-forest model of a suitably bounded branching. To make such a KB dependent only on the TBox, we are going to use the ghost variable $\ghost$ in place of the intended element $\domelemd$.

\begin{definition}\label{def:subtree-consistency}
  Let $\tboxT$ be a $\ZOIQ$-TBox in Scott's normal form and $\aboxS$ be a $\tboxT$-ghost-summary.
  $\aboxS$ is \vocab{$\tboxT$-subtree-consistent} if the $\ZOIQ$-KB $\kbK_{\aboxS,\ghost} \deff (\aboxS, \tboxTloc \cup \tboxT_{\text{aut}} \cup \tboxT_{\text{cnt}})$ is quasi-forest satisfiable, where $\tboxTloc$ is as in \Cref{def:clearing-consistent}~and:\\
  $\bullet$  $\tboxT_{\text{aut}}$ consists~of
      $\{ \ghost \} \dlsubseteq \mathrm{comdsc}(\ind{\tboxT}, \automatonA)$ and $\conceptA \equiv \mathrm{rel}(\ind{\tboxT}, \automatonA_{\stateq, \stateq'})$ for all GCIs $\conceptA \equiv \exists{\automatonA_{\stateq,\stateq'}}.\top$ from $\tboxT$.\\
    $\bullet$ $\tboxT_{\text{cnt}}$ consists of the GCIs $\neg\concept{Root} \dland \conceptA \equiv \neg\concept{Root} \dland ({\geq}{n}\ \roler).\top$ as well as 
    $\{ \ghost \} \dlsubseteq \textstyle\mathrm{desc}\left( \concept{Chld}_{\homot_{\text{ch}}^{\ghost}}^{\roler} \dland \textstyle\bigdland_{\indvo \in \ind{\tboxT}} \concept{Des}_{\homot_{\indvo}^{\ghost} }^{\roler, \indvo}\right),$
   for~the unique concepts $\concept{Chld}_{\homot_{\text{ch}}^{\ghost}}^{\roler}$, $\concept{Des}_{\homot_{\indvo}^{\ghost} }^{\roler, \indvo}$ satisfied by $\ghost$ in $\interI_{\aboxS}$, for all GCIs $\conceptA \equiv ({\geq}{n}\ \roler).\top$ appearing in $\tboxT$.\myqed
\end{definition}

The crucial property concerning the notion of $\tboxT$-subtree-consistency as well as the set \vocab{$\allTconssum$} of all $\tboxT$-subtree-consistent $\tboxT$-ghost-summaries is provided next. 
Its proof relies on the bounds on the concepts from \Cref{lemma:the-main-property-of-A-decorated-quasi-forests}, \Cref{lemma:intended-behaviour-of-A-concepts}, and \Cref{def:IT-r-n-counting-decoration}, as well as the exponential time algorithm for deciding quasi-forest-satisfiability of $\ZOIQ$-KBs from \Cref{lemma:deciding-ZOIQ-ExpTime-complete}.
\begin{lemma}\label{lemma:computing-subtree-consistent-ghost-summaries}
  For a $\ZOIQ$-TBox $\tboxT$ in Scott's normal form, one can decide in time exponential w.r.t. $|\tboxT|$ if a given $\tboxT$-ghost-summary is $\tboxT$-subtree-consistent.
  Moreover, the set~$\allTconssum$ of all $\tboxT$-subtree-consistent $\tboxT$-ghost-summaries has size exponential in $|\tboxT|$ and is computable in time exponential~in~$|\tboxT|$.~\myqed
\end{lemma} 

We now lift the definition of subtree-consistency from ghost summaries to arbitrary $\kbK$-summaries by producing a ghost summary per each individual name $\indva$ mentioned in $\kbK$. 
The idea is simple: we first restrict $\aboxS$ to nominals and the selected name~$\indva$, and then replace $\indva$ with the ghost variable $\ghost$ (we must be a bit more careful if $\indva$ is itself a nominal). This produces a ghost summary, for which the notion of subtree-consistency is already well-defined. The main benefit of testing subtree-consistency of such a summary is that this guarantees the existence of a quasi-forest containing a subtree rooted at $\ghost$ that we can afterwards ``plug in'' to $\indva$ in $\aboxS$  in order to produce a full model~of~$\kbK$~from~$\aboxS$. This is formalised next. 

\begin{definition}\label{def:consistent-summaries}
  Let $\tboxT$ and $\allTconssum$ be as in \Cref{lemma:computing-subtree-consistent-ghost-summaries}.
  For a $\ZOIQ$-KB $\kbK \deff (\aboxA, \tboxT)$ and a $\kbK$-summary $\aboxS$ we say that~$\aboxS$ is \vocab{consistent} if $\aboxS$ is clearing-consistent and for every name $\indva \in \indK$ the \vocab{$(\tboxT, \aboxS, \indva)$-ghost-summary $\aboxS_{\indva}$} belongs to $\allTconssum$. 
   $\aboxS_{\indva}$ is defined as the sum of (i) $\aboxS$ restricted to $\ind{\tboxT}$, (ii) $\aboxS$ restricted to $\ind{\tboxT} \cup \{ \indva \}$ with all occurrences of $\indva$ replaced with $\ghost$, and (iii) $\{ \indva \approx \ghost, \ghost \approx \indva \}$ in case $\indva \in \ind{\tboxT}$.\myqed
\end{definition}

It can be readily verified that all $\kbK$-summaries constructed from elegant models of $\ZOIQ$-KBs $\kbK \deff (\aboxA,\tboxT)$ are consistent. 
For the opposite direction, we define a suitable notion of a  merge in order to ``combine'' a clearing-consistent summary $\aboxS$ with relevant subtrees provided by the subtree-consistency of $(\tboxT, \aboxS, \indva)$-ghost-summaries for all $\indva \in \indK$. The interpretation constructed in this way becomes an elegant model~of~$\kbK$.

\begin{lemma}\label{lemma:zoiq-kb-is-quasi-forest-satisfiable-iff-there-exists-a-consistent-summary}
  A $\ZOIQ$-KB in Scott's normal form is quasi-forest satisfiable iff there exists a consistent $\kbK$-summary.\myqed
\end{lemma}


  \section{Deciding Existence of Quasi-Forest Models}\label{sec:deciding-quasi-forests}

  Based on \Cref{lemma:zoiq-kb-is-quasi-forest-satisfiable-iff-there-exists-a-consistent-summary} and all the other lemmas presented in the previous section, 
  we can finally design an algorithm for deciding whether an input $\ZOIQ$-KB is quasi-forest satisfiable.

  \begin{algorithm}[h]
    \DontPrintSemicolon
    \KwData{A $\ZOIQ$-KB~$\kbK \deff (\aboxA, \tboxT)$.}
    \KwResult{\texttt{True} iff $\kbK$ is quasi-forest satisfiable.}
    \caption{Quasi-Forest Satisfiability in $\ZOIQ$}\label{alg:data-complexity-ZOIQ}
    
    Turn $\tboxT$ into Scott's normal form.
    \tcp*{L.~\ref{lemma:elegant-models-for-scotts-normal-form}.}

    Compute the set $\allTconssum$.
    \tcp*{L.~\ref{lemma:computing-subtree-consistent-ghost-summaries}.}

    \textbf{Guess} a $\kbK$-summary $\aboxS$. 
    \tcp*{L.~\ref{lemma:bounding-sizes-and-numbers-of-summaries}.}

    Return \texttt{False} if $\aboxS$ is clearing-consistent.
    \tcp*{L.~\ref{lemma:deciding-clearing-consistency-is-in-PTime}.}

    \textbf{Foreach} $\indva \in \indK$ \textbf{do} compute the $(\tboxT, \aboxS, \indva)$-ghost- -summary $\aboxS_{\indva}$ and return \texttt{False} if $\aboxS_{\indva} \not\in \allTconssum$. 

    Return \texttt{True}.
  \end{algorithm}

  \begin{lemma}\label{lemma:correctness-of-quasi-forest-sat-algorithm}
    Procedure~\ref{alg:data-complexity-ZOIQ} returns \texttt{True} if and only if the input $\ZOIQ$-KB $\kbK$ has an elegant model. 
    Moreover, there exist polynomial function $\mathrm{p}$ and an exponential function $\mathrm{e}$ for which Procedure~\ref{alg:data-complexity-ZOIQ} can be implemented with a nondeterministic Turing machine of running time bounded, for every input $\kbK \deff (\aboxA, \tboxT)$, by $\mathrm{e}(|\tboxT|)+\mathrm{p}(|\kbK|)+\mathrm{p}(|\kbK|)\cdot\mathrm{e}(|\tboxT|)$.\myqed
  \end{lemma}

  Suppose now that a TBox $\tboxT$ is fixed and only an ABox~$\aboxA$~is given as the input.
  Then the first two steps of Procedure~\ref{alg:data-complexity-ZOIQ} are independent from the input. The same holds for the verification of whether a given ghost summary belongs to the pre-computed set $\allTconssum$. The desired  algorithm is given below. 
  \begin{algorithm}
    \DontPrintSemicolon
    \KwData{An ABox $\aboxA$.}
    \SetKwInput{KwData}{Parameters}
    \KwData{$\ZOIQ$-TBox $\tboxT$ already in Scott's normal form and a pre-computed set $\allTconssum$. }
    \KwResult{\texttt{True} iff $\kbK \deff (\aboxA, \tboxT)$ is quasi-forest satis.}

    \caption{Deciding Quasi-Forest Satisfiability in $\ZOIQ$ w.r.t. Data Complexity}\label{alg:data-complexity-ZOIQ-II}

    \textbf{Guess} a $\kbK$-summary $\aboxS$. 
    \tcp*{$\NP$ in $|\kbK|$, L.~\ref{lemma:bounding-sizes-and-numbers-of-summaries}.}

    Return \texttt{False} if $\aboxS$ is clearing-consistent.\\
  \tcp*{$\PTime$ w.r.t. $|\kbK|{\cdot}|\aboxS|$ by L.~\ref{lemma:deciding-clearing-consistency-is-in-PTime}.}

    \textbf{Foreach} $\indva \in \indK$ \textbf{do} compute the $(\tboxT, \aboxS, \indva)$-ghost- -summary $\aboxS_{\indva}$ and return \texttt{False} if $\aboxS_{\indva} \not\in \allTconssum$. 
    \tcp*{$\PTime$ w.r.t. $|\kbK| \cdot (|\kbK|{+}|\allTconssum|)$}

    Return \texttt{True}.
  \end{algorithm}

  We employ \Cref{lemma:correctness-of-quasi-forest-sat-algorithm}, and based on Procedure~\ref{alg:data-complexity-ZOIQ-II} we get:

  \begin{theorem}\label{thm:main-theorem-about-quasi-forest-sat}
    For every $\ZOIQ$-TBox $\tboxT$, there is an $\NP$ procedure (parametrised by $\tboxT$) that for an input ABox $\aboxA$ decides if the $\ZOIQ$-KB $\kbK \deff (\aboxA, \tboxT)$ is quasi-forest satisfiable.\myqed
  \end{theorem}

  Recall that the maximal decidable fragments of $\ZOIQ$ have the elegant model property (\Cref{lemma:elegant-models-for-scotts-normal-form}). We conclude:
  \begin{theorem}\label{thm:data-complexity-of-ZOIQ}
    The satisfiability problem for $\ZIQ$, $\ZOQ$, and $\ZOI$ is $\NP$-complete w.r.t. the data complexity.\myqed
  \end{theorem}

  For expressive logics from the $\mathcal{SR}$ family, the logical core of OWL2, it is known~\cite[Prop. 5.1]{CalvaneseEO09} that any TBox in $\mathcal{SRIQ}$, $\mathcal{SROQ}$ or $\mathcal{SROI}$ can be rewritten into  $\ZIQ$, $\ZOQ$ or $\ZOI$. Hence, we reprove the following result:

  \begin{corollary}
    The satisfiability problem for $\mathcal{SRIQ}$, $\mathcal{SROQ}$, and $\mathcal{SROI}$ is $\NP$-complete w.r.t. the data complexity.\myqed
  \end{corollary}


\section{Application: Entailment of Rooted Queries}\label{sec:entailment-of-rooted-queries}

We adapt Procedure~\ref{alg:data-complexity-ZOIQ} to derive $\coNExpTime$-completeness of the entailment problem of (unions of) rooted conjunctive queries over $\ZIQ$-KBs, generalising previous results~on $\DL{SHIQ}$ \cite[Thm.~2]{LutzDL08}. 
We focus on the dual problem: ``Given a (union of) rooted {\CQ}s $\queryq$ and $\ZIQ$-KB $\kbK \deff (\aboxA,\tboxT)$, is there a model of $\kbK$ that violates $\queryq$?'' and show its $\NExpTime$-completeness. 
As we work with $\ZIQ$, by \Cref{lemma:elegant-models-for-scotts-normal-form} we can assume that the input KB $\kbK$ is in Scott's normal form, and the intended model $\interI$ violating $\queryq$ is elegant (in particular, is $\rmN$-bounded for an $\rmN$ exponential in $|\tboxT|$).
The crucial observation is that whenever $\queryq$ matches $\interI$, all the elements from the image of a match are at the depth at most $|\queryq|$ (as $\queryq$ has at least one individual name and is connected).
Hence, it suffices to construct an ``initial segment'' of $\interI$ of depth at most $|\queryq|$ and degree at most $\rmN$, and check if (i) $\queryq$ does not match it, and (ii) it can be extended to the full~model~of~$\kbK$.

\begin{definition}\label{def:RCD-forests}
  Let $\mathrm{R}$, $\mathrm{C}$, $\mathrm{D} \in \N$.
  An \vocab{$(\mathrm{R}, \mathrm{C}, \mathrm{D})$-forest $\inter{F}$} is a prefix-closed set of non-empty words from $\ZZ_{\mathrm{R}} {\cdot} (\ZZ_{\mathrm{C}})^+$ of length at most $\mathrm{D}$ ($\ZZ_n$ denotes the set of integers modulo $n$).
  The number~$\mathrm{R}$ indicates the total number of roots of $\inter{F}$, $\mathrm{C}$~denotes the maximal number of children per each element, and $\mathrm{D}$ indicates the maximal depth. 

  Treating $\inter{F}$ as a set of individual names, we define an \vocab{$(\inter{F},\kbK)$-initial segment} of a $\ZOIQ$-KB $\kbK$ as any summary $\aboxS$ of $\kbK \cup \{ \indva \not\approx \indvb \mid \indva,\indvb \in \inter{F}, \indva \neq \indvb \}$~such~that:\\
  $\bullet$ For every $\indva \in \indK$ there is $\indvb \in \inter{F}\cap\N$ with $\indva \approx \indvb$ in $\aboxS$, and for every $\indvb \in \inter{F}\cap\N$ there is  $\indva \in \indK$ with $\indva \approx \indvb$ in $\aboxS$.\\
  $\bullet$ $\left(({=}0\ \role{child}).\top\right)(\indva)$ for all $\indva \in \inter{F}$ that are not leaves of $\inter{F}$.\\
  $\bullet$ $\neg\roler(\indva,\indvb)$ for all role names $\roler$ appearing in $\kbK$ and all $\indva \neq \indvb$ from $\inter{F}$ such that $\indva$ is not a child of $\indvb$ in $\inter{F}$ (or vice-versa).\myqed
\end{definition}

The above conditions are needed to represent a forest $\mathcal{F}$ inside the clearing of the intended models of~$\aboxS$.
The first item of \Cref{def:RCD-forests} guarantees the proper behaviour of roots. 
The second item guarantees that the children of elements in the initial segment are precisely the ones that are explicitly mentioned there.
Finally, the last item ensures that the ``tree-likeness'' of the initial segment is not violated.

\begin{lemma}\label{lemma:about-initial-segments}
  Let $\kbK\deff(\aboxA,\tboxT)$ be a $\ZIQ$-KB in Scott's normal form and let $\queryq \deff \textstyle \bigvee_i \queryq_i$ be a union of rooted {\CQ}s.
  We have that $\kbK \not\models \queryq$ if and only if there is an $(\inter{F},\kbK)$-initial segment~$\aboxS$ such that (i)~$\inter{F}$~is an $(\mathrm{R}, \mathrm{C}, \mathrm{D})$-forest for an $\mathrm{R}$ bounded by $|\indK|$, $\mathrm{C}$ exponential in $|\tboxT|$,~and $\mathrm{D}$ bounded by~$|\queryq|$, 
  (ii) $(\aboxS, \tboxT)$ is quasi-forest satisfiable, 
  and (iii)~$\interI_\aboxS \not\models \queryq$.\myqed
\end{lemma}

\noindent Note that the forest $\inter{F}$ described above is exponential w.r.t.~$|\kbK|$, and thus the initial segment $\aboxS$ can be ``guessed'' in~$\complexityclass{NExp}$. 

\begin{algorithm}
  \DontPrintSemicolon
  \KwData{A $\ZIQ$-KB~$\kbK \deff (\aboxA, \tboxT)$ and a union $\queryq \deff \textstyle\bigvee_{i=1}^{n} \queryq_i$ of rooted conjunctive queries.}
  \KwResult{\texttt{True} if and only if $\kbK \not\models \queryq$.}
  \caption{Rooted Query Entailment in $\ZIQ$}\label{alg:rooted-entailment-for-ZOIQ}
  
  Turn $\tboxT$ into Scott's normal form.

  \textbf{Guess} $(\inter{F},\kbK)$-initial segment $\aboxS$ of $\kbK$ as in \Cref{lemma:about-initial-segments}. 

  \textbf{Foreach} $1 \leq i \leq n$ return \texttt{False} if $\interI_{\aboxS} \models \queryq_i$.

  Use Procedure~\ref{alg:data-complexity-ZOIQ} to verify that $(\aboxS, \tboxT)$ is quasi-forest satisfiable and return \texttt{True} if this is indeed the case. 
\end{algorithm}

\begin{lemma}\label{lemma:correctness-of-quasi-forest-sat-rooted-algorithm}
  Procedure~\ref{alg:rooted-entailment-for-ZOIQ} returns \texttt{True} if and only if the input $\ZOIQ$-KB $\kbK$ does not entail the input query $\queryq$. 
  Moreover, Procedure~\ref{alg:rooted-entailment-for-ZOIQ} can can be implemented with a nondeterministic Turing machine of running time bounded by some function bounded exponentially w.r.t. $|\queryq|$ and $|\kbK|$.\myqed
\end{lemma}

Based on \Cref{lemma:correctness-of-quasi-forest-sat-rooted-algorithm} we can conclude:  
\begin{theorem}\label{thm:rooted-entailment-for-ZIQ-is-coNExpTime-complete}
  The entailment problem for unions of rooted {\CQ}s over $\ZIQ$-KBs is $\coNExpTime$-complete.\myqed
\end{theorem}

The $\coNExpTime$-hardness of rooted entailment holds already for $\ALCI$~\cite[Thm.~1]{LutzIJCAR08}. Interestingly enough, we establish [Appendix L] a novel lower bound for $\ALC\Self$.
\begin{theorem}\label{thm:rooted-entailment-for-ALCSelf-is-coNExpTime-hard}
  The rooted conjunctive query entailment problem is $\coNExpTime$-hard for $\ALC\Self$ (thus also for $\ZIQ$).\myqed
\end{theorem}

\section*{Acknowledgements}
This work was supported by the ERC Grant No.~771779 (\href{https://iccl.inf.tu-dresden.de/web/DeciGUT/en}{DeciGUT}).
Full version of this paper is available on ArXiV.

\bibliographystyle{named}
\bibliography{references}

\end{document}